\documentclass{jpp}
\usepackage{graphicx}
\usepackage[utf8]{inputenc}
\usepackage[T1]{fontenc}
\usepackage{amsmath}
\usepackage{xcolor}
\usepackage{changes}

\shorttitle{Temperature inversion in a non-homogeneous plasma}
\shortauthor{L. Barbieri, E. Papini, P. Di Cintio, S. Landi, A. Verdini, L. Casetti}
\title{Temperature inversion in a confined plasma atmosphere: coarse-grained effect of temperature fluctuations at its base}
\author{Luca Barbieri\aff{1,2,3,}
  \corresp{\email{luca.barbieri@unifi.it}},
  Emanuele Papini\aff{4},
  Pierfrancesco Di Cintio\aff{2,3,5},\\
  Simone Landi\aff{1,2},
  Andrea Verdini\aff{1,2},
Lapo Casetti\aff{1,2,3}
 }
\affiliation{\aff{1}Dipartimento di Fisica e Astronomia, Universit\`a di Firenze, via Giovanni Sansone 1, Sesto Fiorentino, I-50019, Italy,
\aff{2}INAF - Osservatorio Astrofisico di Arcetri, Largo Enrico Fermi 5, Firenze, I-50125, Italy
\aff{3}INFN - Sezione di Firenze, via Giovanni Sansone 1, Sesto Fiorentino, I-50019, Italy
\aff{4}INAF - Istituto di Astrofisica e Planetologia Spaziali, via del Fosso Cavaliere 100, Roma, I-00133, Italy
\aff{5}Istituto dei Sistemi Complessi, Consiglio Nazionale delle Ricerche (ISC-CNR), via Madonna del piano 10, Sesto Fiorentino, I-50019, Italy
}
\begin{document}
\maketitle
\begin{abstract}
Prompted by the relevant problem of temperature inversion (i.e. gradient of density anti-correlated 
to the gradient of temperature) in astrophysics, we introduce a novel method to model a gravitationally confined multi-component collisionless plasma in contact with a fluctuating thermal boundary. We focus on systems with anti-correlated (inverted)  density and temperature profiles, with applications to solar physics. The dynamics of the plasma is analytically described via the coupling of an appropriated coarse-grained distribution function and temporally coarse-grained Vlasov dynamics. We derive a stationary solution of the system and predict the inverted density and temperature profiles of the two-species for scenarios relevant for the corona.
We validate our method by comparing the analytical results with kinetic numerical simulations of the plasma dynamics in the context of the two-species Hamiltonian mean-field model (HMF). Finally, we apply our theoretical framework to the problem of the temperature inversion in the solar corona obtaining density and temperature profiles in remarkably good agreement with the observations.
\end{abstract}
\maketitle
\section{Introduction}
Stationary states out of thermal equilibrium occur in nature and often exhibit unexpected properties.
An example is given by the so-called inverted temperature-density profiles, or temperature inversion, that corresponds to an increase in temperature while the density is decreasing. Numerical simulations have shown how temperature inversion can be achieved in non-equilibrium stationary states, usually referred to as quasi-stationary states, attained after processes of violent relaxation (see e.g.\ \citealt{Lynden-Bell1967,1998ApJ...500..120K,2008PhRvL.100d0604L,ewart_brown_adkins_schekochihin_2022}, see also \citealt{Barbieri2022} and references therein), of many-body systems with long-range inter-particle potentials (see \citealt{2013MNRAS.431.3177D,Campa2014} and references therein), brought out of thermal equilibrium by energy injection (\citealt{Casetti2014,Teles-Casetti2015,Gupta_2016}).\\
\indent Temperature inversion occurs in several astrophysical systems such as, for example, filaments in molecular clouds (\citealt{Arzoumanian:aa2011,TociGalli:mnras2015,DiCintio2017}) the Io plasma torus around Jupiter (\citealt{MeyerVernetMoncuquetHoang:icarus1995}) and the hot gas in galaxy clusters (\citealt{BaldiEtAl:apj2007,WiseMcNamaraMurray:apj2004}). The most relevant and widely studied case is represented by the atmosphere of the Sun (\citealt{GolubPasachoff:book}). The problem of temperature inversion in the atmosphere of the Sun has been approached in several ways (see \citealt{Klimchuk_2006,2012coronalheating} and references therein) but remains far from being completely solved.\\
\indent Among all the existing theoretical models, the one that is relevant for the present work is the kinetic approach introduced by \cite{Scudder1992a,Scudder1992b} and dubbed by the author ``velocity filtration''. In this approach the solar atmosphere is treated as a system of non-interacting particles under the action of the Sun's gravity. Thanks to the conservation of energy, only particles with a sufficient amount of kinetic energy can climb the gravity well and reach a given altitude, thus forming the atmosphere (for this reason the mechanism was dubbed ``velocity filtration''). If the particle's velocity distribution functions (VDFs) at the base of the atmosphere are thermal, then the stationary configuration is the standard isothermal stratified atmosphere (\citealt{2005psci.book.....A}), with a density profile exponentially decreasing with altitude and a constant temperature all over the atmosphere. If, as in the original Scudder model, the velocity distribution functions are suprathermal (e.g., they are given by a Kappa distribution, see \citealt{lazar2021kappa}), then the temperature profile is a growing function of the atmosphere’s height, thus resulting in stationary anti-correlated density and temperature profiles. In Scudder formalism, however, the chromospheric velocity distribution functions are assumed fixed and suprathermal, even if the chromosphere is highly collisional, posing the problem on how such distributions can be sustained in such an environment. Moreover, modelling the chromosphere with a fixed Kappa function (as in \cite{Scudder1992a,Scudder1992b}) produces a linear increase of temperature rather than a transition region and corona, as observed in the solar atmosphere. To overcome the difficulty of sustaining suprathermal populations in the strongly collisional chromosphere, mechanisms related to turbulence and wave-particles interactions \citep[e.~g.][]{Parker1958} or statistical processes based on Levy's flights \citep{Collier1993} have been often evoked. In the present work, as well as in \cite{barbieri2023temperature}, we want to overcome this difficulty by modelling the chromosphere as fully collisional so that it can be schematised at every instant of time as a thermal boundary. We then ask if rapid heating events, modelled as sudden temperature increments of the thermal boundary, are able to produce suprathermal distribution functions. More precisely, in \cite{barbieri2023temperature} it has been shown that rapid, intermittent and short-lived heating events are able to drive the coronal collisionless plasma towards a non-equilibrium stationary state characterised by inverted temperature-density profiles, with a transition region and a corona similar to those observed in the solar atmosphere. As pointed out in \cite{barbieri2023temperature} the idea behind studying the role of temperature fluctuations in the chromosphere lies in the fact that while the VDFs of the chromospheric plasma are likely to be thermal due to collisionality, the chromosphere is a very dynamic environment, showing fine-scale structures down to instrumental resolution (see \cite{Molnar_2019,Ermolli:2022um}), hence its temperature is expected to fluctuate in space and time \citep{Peter:2014uz,Hansteen:2014us}.\\
In this paper we investigate this matter further, by analysing the problem in details from the point of view of kinetic plasma physics. We introduce a novel temporal coarse-graining technique to model the physics of the non-equilibrium plasma dynamics described above and in \cite{barbieri2023temperature}. Here we show how the dynamics can be described by a set of two coupled Vlasov equations for the temporal coarse-graining version of the distribution functions of the two species (here electrons and protons). This system of equations is coupled to effective coarse-grained distributions at its boundary describing the temperature fluctuations of the chromosphere. We stress the fact that, at variance with \cite{Scudder1992a,Scudder1992b}, we do not consider non-interacting particles, but rather we explicitly take into account their mutual electrostatic interaction, although only at the mean-field level.   
We derive analytical expressions for the coarse-grained distribution functions of the two species in the non-equilibrium stationary configuration. 
These distribution functions exhibit suprathermal high-energy tails because of the dynamics induced by the temperature fluctuations of the thermal boundary, that is, they are self-consistently produced by our modelling of the chromosphere. 
Of course, extra parameters are introduced to specify the distribution of temperature fluctuations, but no fine-tuning of these parameters is necessary to produce the temperature inversion and for allowing the use of a coarse-graining approach. The only requirement is a fluctuation time scale that is smaller than the relaxation time scale of the system. 
As a consequence, our coarse-graining formalism lets us interpret our results in terms of Scudder's velocity filtration mechanism, but without imposing a priori suprathermal tails. 
We note that suprathermal tails in the distribution functions can be obtained even in stationary states of an isolated collisionless plasma as a consequence of the dynamical phase-mixing in the single-particle phase-space (see \citealt{2023arXiv230403715E} and references therein).\\
\indent The paper is structured as follows. In Section 2 we present the model used to describe the collisionless plasma atmosphere in contact with the fluctuating thermal boundary. In Section 3 we present the temporal coarse-graining approach describing the plasma dynamics and we obtain the coarse-grained version of the distribution functions of the two species in the non-equilibrium stationary configuration. In Section 4 we test the theory against kinetic numerical simulations and we discuss its limits of applicability. In Section 5 we apply our method to the specific case of the atmosphere of the Sun and discuss the results, also comparing our findings with the observed density and temperature profiles of \citealt{observedtemperature}. Moreover, we discuss some observational evidences to support our result and the physical limits of modelling the coronal plasma neglecting Coulomb collisions.
Finally, in Section 6 we summarise and discuss the future perspectives of this study.
\section{The two-component gravitationally bound plasma model}\label{sec2}
\begin{figure}
    \centering
    \includegraphics[width=0.9\columnwidth]{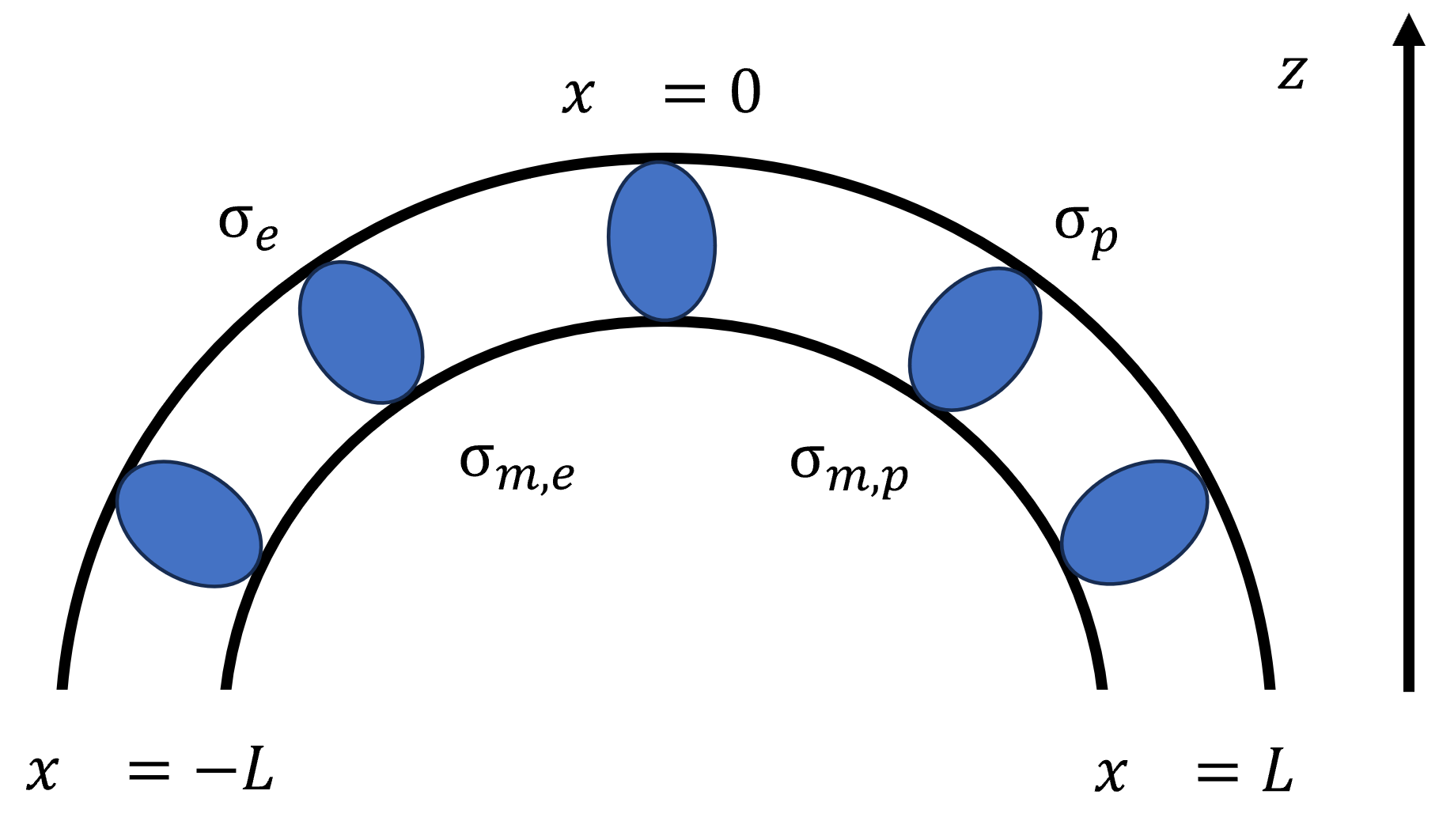}
    \caption{Sketch of the two-component plasma loop model. The vertical axis $z$ is the altitude in the atmosphere; $x$ is the curvilinear abscissa of the loop. $\sigma_{m,\alpha}$,$\sigma_{\alpha}$ with $\alpha=\{e,p\}$ are respectively the surface mass density and the surface charge density of the species $\alpha$.}
    \label{fig:coronalloop}
\end{figure}

We model a gravitationally bound plasma, focusing on geometrically confined plasma structures, with specific reference to the coronal loops that are common in the Sun's atmosphere (see e.g. \citealt{2005psci.book.....A}). We model the loop as a semicircular tube of length $2L$ and section $S$, where the charge distribution is discretized onto $N_e$ sections with surface number density $n_S$, surface charge density $\sigma_e=-e n_S$ and surface mass density $\sigma_{m,e}=m_e n_s$; and $N_p$ sections with surface charge density $\sigma_p=e n_S$ and surface mass density $\sigma_{m,p}=m_p n_S$, representing the electron and proton components, respectively. We assume that $N_e=N_p=2N$ in order to ensure global charge neutrality. A scheme of a two-component loop plasma model is sketched in Figure \ref{fig:coronalloop}. We now introduce the following assumptions:
\begin{enumerate}
    \item Particles are subject to an external force field pointing towards the loop's end, representing the (uniform) Sun's gravity plus the Pannekoek-Rosseland electrostatic field (\citealt{Pannekoek_1922,Rosseland_1924}, see also \citealt{belmont2013collisionless} for an extended review). 
    \item The symmetry is cylindrical along the loop-following coordinate $x$. 
    \item Electrostatic self-interactions are modelled by truncating to the first order Fourier expansion as in the so-called Hamiltonian Mean-Field models (hereafter HMF, see \citealt{AntoniRuffo:pre1995,GiachettiCasetti:jstat2019,Chavanis2005}).
    \item Every observable is symmetric with respect to the top of the loop \footnote{The physical reason behind this idea is that we can schematise the chromosphere (i.e. the thermal boundary) with its average properties (in space and time) at both feet of the Loop. Combining this approximation with the symmetry of the Loop structure with respect to the top, we can impose the central symmetry during the whole dynamics.}. 
    As a consequence, if the top of the loop coincides with $x = 0$, all the functions of the canonical coordinates are centrally symmetric with respect to the origin of the two-dimensional single-particle phase space \footnote{The practical realisation of that will be presented in Section 2.1 }.  
\end{enumerate}
The first two assumptions imply that the total energy of the system is
\begin{equation}\label{1Dplasma}
    H=S\biggl(\sum_{j=1}^{2N}\sum_{\alpha \in \{e,p\}} \frac{1}{2}\sigma_{m,\alpha} \dot{x}^2_{j,\alpha} + E_{el} + U_{g} \biggl),
\end{equation}
where $\alpha \in \{e,p\}$ denotes the species (here electrons or protons), $g={G M_{\odot}}/{R_{\odot}^2}$ is the
gravitational field at the surface of the Sun, $x$ is the spatial coordinate (i.e. the curvilinear abscissa of the loop). The total electrostatic energy per section  $E_{el}$ reads
\begin{equation}\label{Electrostaticenergy}
\begin{gathered}
    E_{el}=\frac{1}{2}\int_{-L}^{L} d x \rho(x)\phi(x),
\end{gathered}
\end{equation}
where the charge density 
\begin{equation}\label{chargedens}
\rho(x)= \sum_{j=1}^{2N}\sum_{\alpha \in \{e,p\}} \sigma_{\alpha} \delta (x-x_{j,\alpha}), 
\end{equation} 
 is related to the potential $\phi$ by the Poisson equation 
 \begin{equation}\label{poissoneq}
  \frac{\partial^2 \phi}{\partial x^2}=-4\pi\rho.
 \end{equation}
 $U_g$ is the total external potential per section and reads
 \begin{equation}\label{Gravitationalenergy}
    U_g=\frac{gL\sum_{\alpha \in \{e,p\}} \sigma_{m,\alpha}}{\pi}\sum_{j=1}^{2N}\sum_{\alpha \in \{e,p\}} \cos{\biggl(\frac{\pi x_{j,\alpha}}{2L}\biggl)}.
\end{equation}
In order to use the HMF modelization we Fourier analyze the electrostatic energy $E_{el}$. Hereafter we will use the following definition for the Fourier transform
\begin{equation}\label{FourierModes}
    f(x)=\sum_{n\in \mathbb{Z}} f_n e^{i\frac{\pi n x}{L}}, \quad\quad f_n=\frac{1}{2L}\int_{-L}^L dx f(x) e^{-i\frac{\pi n x}{L}}.
\end{equation}
We then define the charge imbalances as 
\begin{equation}\label{Chargeimbalances}
\textbf{Q}_{n}=(Q_{n,x},Q_{n,y})=\sum_{\alpha \in \{e,p\}}\mathrm{sign}{(\sigma_{\alpha})}\textbf{q}_{n,\alpha},
\end{equation}
where the vectors $\textbf{q}_{n,\alpha}$ are given by
\begin{equation}\label{Stratificationparameters}
    \textbf{q}_{n,\alpha}=\frac{1}{2N}\sum_{j=1}^{2N} \biggl[ \cos{\biggl(\frac{\pi n x_{j,\alpha}}{L}\biggl)},\sin{\biggl(\frac{\pi n x_{j,\alpha}}{L}\biggl)}\biggl]. 
\end{equation}
Performing the Fourier transform \eqref{FourierModes} of the Poisson equation (\ref{poissoneq}) and applying Eqs. (\ref{Chargeimbalances}-\ref{Stratificationparameters}) we obtain the modes of the potential-density pair as
\begin{equation}\label{ChargeinFourier}
\phi_n=\frac{4L^2}{\pi n^2}\rho_n, \quad 
\rho_n=-\frac{\sigma_e N}{L}(q_{n,x}-i q_{n,y}) \quad \forall n \neq 0 \quad \rho_0=0.
\end{equation}
 Since each density mode $\rho_n$ defined above is related to a specific $\textbf{Q}_n$, any vanishing density mode $\rho_n$ corresponds to a zero charge imbalance. This can be also seen directly from Eq.\ \eqref{Chargeimbalances} and Eq.\ \eqref{Stratificationparameters}. Indeed, if most of the $2N$ particles of a given species are symmetrically concentrated at the bottom of the loop, then $\textbf{q}_{n,\alpha} \approx -1$. Indeed, if they are uniformly distributed then $\textbf{q}_{n,\alpha} \approx 0$ ($\textbf{q}_{n,\alpha} \approx +1$ corresponds to a situation where most of the particles are concentrated at the top of the loop). As a consequence, if their difference $\textbf{Q}_n$ does not vanish,  there is a charge imbalance in the system.\\
\indent From now on we will refer to the quantities $\textbf{q}_{n,\alpha}$ as the stratification parameters. By performing the Fourier expansion \eqref{FourierModes} of the charge density \eqref{chargedens} and of the electrostatic potential $\phi(x)$ and by using Eqs. (\ref{Chargeimbalances}-\ref{ChargeinFourier}), after some algebra we get the decomposition in Fourier modes of the electrostatic energy per section \eqref{Electrostaticenergy} as
\begin{equation}\label{Electrostaticenergyallmodes}
    E_{el}=\frac{8L (N\sigma_e)^2}{\pi}\sum_{n=1}^{+\infty} \frac{||\textbf{Q}_n||^2}{n^2}.
\end{equation}
The proportionality to $(N\sigma_e)^2$ arises from the fact that each particle of the system feels the electrostatic forces due to all the rest of the particles. Its proportionality to $L$ means the larger the size of the system is, the greater is $E_{el}$. The electrostatic energy $E_{el}$ depends on all the charge imbalances $||\textbf{Q}_n||^2$ scaled by $n^2$, that is, large-scale modes give larger contribution to the electrostatic energy compared to the small-scale ones. Using Eq. \eqref{Electrostaticenergyallmodes} combined with the total energy \eqref{1Dplasma} we derive the equations of motion in the form
\begin{equation}
m_{\alpha} \ddot{x}_{j,\alpha}=eE(x_{j,\alpha})+g \frac{\sum_{\alpha \in \{e,p\}}m_{\alpha}}{2}\sin{\biggl(\frac{\pi x_{j,\alpha}}{2L}\biggl)}, 
\end{equation}
where the electric field $E$ decomposed in Fourier modes is
\begin{equation}
E(x)=8~\mathrm{sign}{(e_{\alpha})}e_{\alpha}n_SN\sum_{n=1}^{+\infty} \biggl[\frac{Q_{n,x}}{n}\sin{\biggl(\frac{\pi n x}{L}\biggl)}-\frac{Q_{n,y}}{n}\cos{\biggl(\frac{\pi n x}{L}\biggl)}\biggl].
\end{equation}
We now implement the HMF assumption by truncating the Fourier expansion at the first mode. The equations above become
\begin{equation}\label{HMF2S1}
m_{\alpha} \ddot{x}_{j,\alpha}=e_{\alpha}E(x_{j,\alpha})+ g \frac{\sum_{\alpha \in \{e,p\}}m_{\alpha}}{2}\sin{\biggl(\frac{\pi x_{j,\alpha}}{2L}\biggl)},
\end{equation}
and the electric field becomes
\begin{equation}\label{ElectricfieldHMF2S}
E(x)=8~\mathrm{sign}(e_{\alpha})e_{\alpha}n_SN \biggl[Q_{x}\sin{\biggl(\frac{\pi x}{L}\biggl)}-Q_{y}\cos{\biggl(\frac{\pi x}{L}\biggl)}\biggl].
\end{equation}
In the equations above $\textbf{q}_{\alpha}=\textbf{q}_{1,\alpha}$ are the first mode (large scales) stratification parameters and $Q_x=Q_{1,x}$, $Q_y=Q_{1,y}$ are the first mode charge imbalance components. From now on, unless explicitly stated, we will refer to $\textbf{q}_{\alpha}$ and $\textbf{Q}$ respectively as the stratification parameters and the charge imbalance. As one can see, the HMF assumption reduces the computational cost of the electrostatic interactions from $N^2$ to $N$.\\
\indent Let us now comment further on the physics behind the expression of the self-electrostatic interactions. To do so, we use again the electrostatic energy in terms of its Fourier modes \eqref{Electrostaticenergyallmodes} truncating the expression at the first mode. After some elementary algebra we get
\begin{equation}
\begin{gathered}
E_{el}=E_0\sum_{j,k=1}^{2N}\sum_{\alpha,\beta \in \{e,p\}}\tilde{E}_{\alpha,\beta}\textbf{q}_{j,\alpha}\cdot\textbf{q}_{k,\beta},
\end{gathered} 
\end{equation}
where
\begin{equation}
\textbf{q}_{j,\alpha}=\biggl[\cos{\biggl(\frac{\pi x_{j,\alpha}}{L}\biggl)},\sin{\biggl(\frac{\pi x_{j,\alpha}}{L}\biggl)}\biggl],
\end{equation}
$\tilde{E}_{\alpha,\beta}=\mathrm{sign}(e_{\alpha})\,\mathrm{sign}(e_{\beta})$, and $\quad E_0={4\pi^{-1} L e^2n_S^2N}$. From these expressions it is apparent that the interactions among particles of the same species are described by an antiferromagnetic HMF term while the interactions among particles of different species are, conversely, described by a ferromagnetic HMF term. The antiferromagnetic HMF tends to increase the angle between $\textbf{q}_{j,\alpha}$ and $\textbf{q}_{k,\beta}$ in order to minimise the energy. In terms of physical quantities in the loop this interaction tends to increase the physical distance $x_{j,\alpha}-x_{k,\alpha}$ between two particles of the same species $\alpha$ up to the value $L$ and so such term mimics the electrostatic repulsion between charges of the same sign. The ferromagnetic HMF tends to minimise the energy by decreasing the distance $x_{j,\alpha}-x_{k,\beta}$ between two particles of different species $\alpha,\beta$ up to zero and this mimics the electrostatic attraction between charges of the different sign.

In the following, we consider this plasma model in thermal contact with a thermal boundary at temperature $T_0$ at its boundary (the base of the plasma atmosphere).

\subsection{Normalization and central symmetry assumption}
To study the dynamics defined by Equations (\ref{HMF2S1}), we define the following units for velocity, mass and length
\begin{equation}\label{Setsofunits}
\begin{gathered}
    v_0=\sqrt{\frac{k_B T_{0}}{m_e}}, \quad m_0=m_e, \quad  L_0=\frac{L}{\pi}.
\end{gathered}
\end{equation}
The choice of units given above implies that the unit of energy is $E_0 = k_B T_0$, where $T_0$ is the unit of temperature and $k_B$ is the Boltzmann constant. For the specific case of the Sun the unit of temperature $T_0$ is set equal to the reference temperature of the thermal boundary (chromosphere), that is $T_0=10^4$ K. 
The equations of motion are expressed in dimensionless form as
\begin{equation}\label{HMF2Sdimensionless}
        M_{\alpha} \ddot{\theta}_{j,\alpha}= \mathrm{sign}{(e_{\alpha})} C E(\theta_{j,\alpha})+\tilde{F}(\theta_{j,\alpha}),
\end{equation}
where the expressions for the external forces and the electrostatic field are
\begin{equation}\label{HMF2Sdimensionless2}
        \tilde{F}(\theta_{j,\alpha})= \tilde{g} \sin{\biggl(\frac{\theta_{j,\alpha}}{2}\biggl)}, \quad E(\theta)=Q_{x}\sin{\theta}-Q_{y}\cos{\theta},\\
\end{equation}
and the charge imbalance and stratification parameters are given by
\begin{equation}\label{HMF2Sdimensionless3}
        \textbf{Q}=(Q_x,Q_y)=\sum_{\alpha \in \{p,e\}} \mathrm{sign}{(e_{\alpha})}\textbf{q}_{\alpha},\quad \textbf{q}_{\alpha}=\frac{1}{2N}\sum_{j=1}^{2N}(\cos{(\theta_{j,\alpha})},\sin{(\theta_{j,\alpha})}).
\end{equation}
In the equations above $M_{\alpha}$ equals $M_p ={m_p}/{m_e}$ for the protons; and $M_e = 1$ for the electrons, while $\theta$ is the dimensionless spatial coordinate. The dynamics is therefore fully identified by the three parameters
\begin{equation}
        M=\frac{m_p}{m_e}, \quad
        C=\frac{8 e^2 L^2 n_0}{\pi k_B T_0}, \quad
        \tilde{g}=\frac{g L (m_p+m_e)}{2 \pi k_B T_{0}}, 
\end{equation}
where $n_0=Nn_S/L$ is the average density of a given species. The quantities $C$ and $\tilde{g}$ measure the strength of the electrostatic interaction and of the external field in units of thermal energy, respectively. Hereafter, unless explicitly stated, we will use dimensionless quantities in all the equations and for all the quantities to be plotted in the figures.\\
\indent Assuming that all quantities are symmetric respect to the top of the Loop, in terms of the particles phase-space coordinates, corresponds to imposing the following symmetry rules
\begin{equation}\label{centralsymmetryassumption}
        \theta_{j+N,\alpha}=-\theta_{j,\alpha}, \quad
        \dot{\theta}_{j+N,\alpha}=-\dot{\theta}_{j,\alpha}; \quad \forall j=1 \dots N,
\end{equation}
where the first $j=1 \dots N$ particles populate one half of the loop $\theta_{j,\alpha} \in [-\pi,0]$ and the remaining particles populate the other half. With such an assumption the system of equations \eqref{HMF2Sdimensionless} can be reduced to a system of equations for only the first half of the loop as
\begin{equation}\label{HMF2SdimensionlessCS}
    M_{\alpha} \ddot{\theta}_{j,\alpha}= \mathrm{sign}{(e_{\alpha})} C E(\theta_{j,\alpha})+\tilde{F}(\theta_{j,\alpha}),
\end{equation}
where the Eqs. \eqref{HMF2Sdimensionless2} become
\begin{equation}
    \tilde{F}(\theta_{j,\alpha})= \tilde{g} \sin{\biggl(\frac{\theta_{j,\alpha}}{2}\biggl)},\quad E(\theta)=Q\sin{\theta},
\end{equation}
and the Eqs. \eqref{HMF2Sdimensionless3} become 
\begin{equation}
    Q=\sum_{\alpha \in \{p,e\}} \mathrm{sign}{(e_{\alpha})}q_{\alpha},\quad q_{\alpha}=\frac{1}{N}\sum_{j=1}^{N}\cos{(\theta_{j,\alpha})}.
\end{equation}
The positions and velocities of particles in the other half of the loop are then determined using Eq. \eqref{centralsymmetryassumption}.
\subsection{Vlasov dynamics and thermal equilibrium solution}
In the continuum limit, in terms of phase-space distribution functions the mean-field dynamics of the normalized plasma model given by Eqs.\ \eqref{HMF2SdimensionlessCS} is governed by a system of two Vlasov equations
\begin{equation}\label{Vlasovequations}
    \frac{\partial f_\alpha}{\partial t}+\frac{p}{M_{\alpha}}\frac{\partial f_{\alpha}}{\partial \theta}+F_{\alpha}[f_{\alpha}]\frac{\partial f_{\alpha}}{\partial p}=0, \quad  F_\alpha=-\frac{\partial H_{\alpha}}{\partial \theta},
\end{equation}
where $f_\alpha$ are the single-particle distribution functions for both species, and $H_{\alpha}$ are the mean-field Hamiltonians
\begin{equation}\label{Mean-field-Hamiltonians}
        H_\alpha=\frac{p^2}{2M_\alpha}+\mathrm{sign}{(e_\alpha)}C \phi(\theta)+2\tilde{g}\cos{\biggl(\frac{\theta}{2}\biggl)}; \quad
        \phi(\theta)=Q\cos{\theta},
\end{equation}
where
\begin{equation}\label{qVlasov}
        Q=\sum_{\alpha \in \{e,p\}} \mathrm{sign}{(e_{\alpha})} q_{\alpha}, \quad
        q_{\alpha}=\int_{-\pi}^{\pi} d\theta \int_{-\infty}^{\infty}dp \cos{\theta} f_{\alpha}(\theta,p).
\end{equation}
In virtue of the Jeans' theorem (see e.g. \cite{nicholson1983introduction}), all stationary solution of the system \eqref{Vlasovequations} are functions the sole mean-field Hamiltonians \eqref{Mean-field-Hamiltonians}. Technically speaking, in general, the stationary solution of a Vlasov equation is function of all the constants of motion. Here, since our model is one-dimensional the only independent constants of motion are the mean-field Hamiltonians $H_\alpha$. Among all the solutions of the system above, the thermal equilibrium one has the following expression
\begin{equation}\label{ThermalequilibriumE}
\begin{gathered}
    f_{\alpha}(\theta,p)=\frac{e^{-\frac{H_{\alpha}}{T}}}{Z_{\alpha}}, \quad Z_{\alpha}=\int_{-\pi}^{\pi}d \theta \int_{-\infty}^{+\infty}dp\, e^{-\frac{H_{\alpha}}{T}},
\end{gathered}
\end{equation}
where $f_{\alpha}$ is normalized to 1 for both species. By combining the above equation with Eq.\eqref{qVlasov}, we obtain
\begin{equation}\label{Thermalesystem}
    \begin{gathered}
    Q=\sum_{\alpha \in \{e,p\}} \mathrm{sign}{(e_{\alpha})} \int_{-\infty}^{+\infty} dp \int_{-\pi}^{\pi} d\theta \cos{\theta} \frac{e^{-\frac{H_{\alpha}[Q]}{T}}}{\int_{-\infty}^{+\infty} dp\int_{-\pi}^{\pi}d\theta e^{-\frac{H_{\alpha}[Q]}{T}}},
    \end{gathered}
\end{equation}
and integration with respect to $p$ gives
\begin{equation}\label{Thermalsystem2}
Q=\int_{-\pi}^{\pi}d\theta \cos{\theta}\frac{e^{-\frac{CQ\cos{\theta}+\tilde{g}\cos{\bigl(\frac{\theta}{2}\bigl)}}{T}}}{\int_{-\pi}^{\pi}d\theta e^{-\frac{CQ\cos{\theta}+\tilde{g}\cos{\bigl(\frac{\theta}{2}\bigl)}}{T}}}-\int_{-\pi}^{\pi}d\theta \cos{\theta}\frac{e^{-\frac{-CQ\cos{\theta}+\tilde{g}\cos{\bigl(\frac{\theta}{2}\bigl)}}{T}}}{\int_{-\pi}^{\pi}d\theta e^{-\frac{-CQ\cos{\theta}+\tilde{g}\cos{\bigl(\frac{\theta}{2}\bigl)}}{T}}}   
\end{equation}
whose solution is $Q=0$ (and so $\phi=0$). Therefore, the equilibrium solution of a two-component plasma model, where the two species are subject to the same external potential, requires a vanishing self-electrostatic potential, (i.e., $\phi=0$). The thermal solution for the plasma atmosphere model is the so-called isothermal atmosphere given by
\begin{equation}\label{thermalequilibriumsolution}
    f_{\alpha}(\theta,p)=\frac{e^{-\frac{\tilde{H}_{\alpha}}{T}}}{Z_{\alpha}}, \quad
    Z_{\alpha}=\int_{-\pi}^{\pi} d\theta \int_{-\infty}^{+\infty} dp\,  e^{-\frac{\tilde{H}_{\alpha}}{T}}\,
\end{equation}
where $\tilde{H}_{\alpha}$ are the mean-field Hamiltonians \eqref{Mean-field-Hamiltonians} with $\phi=0$, that is
\begin{equation}\label{Stationarymeanfieldhamiltonians}
    \tilde{H}_\alpha=\frac{p^2}{2M_\alpha}+2\tilde{g}\cos{\biggl(\frac{\theta}{2}\biggl)}.
\end{equation}
The knowledge of the distribution functions allows us to compute the number densities $n_\alpha$ and the temperatures profiles $T_\alpha$ using the standard kinetic definitions (see e.g. \cite{nicholson1983introduction}). For the number densities we get
\begin{equation}\label{Eqdensity}
    n_{\alpha}(\theta)=\frac{e^{-\frac{2\tilde{g}}{T}\cos{\bigl(\frac{\theta}{2}}\bigl)}}{\int_{-\pi}^{\pi}d\theta e^{-\frac{2\tilde{g}}{T}\cos{\bigl(\frac{\theta}{2}}\bigl)}},
\end{equation}
and for the kinetic temperatures
\begin{equation}\label{Eqtemperature}
    T_{\alpha}(\theta)=T.
\end{equation}
\section{Temporal coarse-graining}\label{sec3}
Starting from a thermal equilibrium state \eqref{thermalequilibriumsolution} with temperature $T_0=1$, the out-of-equilibrium state is induced by introducing fluctuations in the value of the temperature of the thermal boundary. In practice, we first increase the temperature from the initial equilibrium value $T_0$ up to $T>T_0$ (sorted from a probability distribution $\gamma(T)$) for a finite fixed time interval $\tau$ \footnote{The choice of using a fixed $\tau$ is made here for the sake of simplicity. We note, however, that using a $\tau$ value drawn by a probability distribution would not change the results.}
, of length smaller than the relaxation time $t_R$ to equilibrium at temperature $T$. After a time $\tau$ the temperature of the thermal boundary is reverted to $T_0$ for a finite time interval $t_w$ sorted from a probability distribution $\eta(t_w)$, again much smaller than the relaxation time $t_R^\prime$ back to $T_0$. Iterating this procedure prevents the system to relax to a thermal equilibrium at either $T_0$ or $T$. 
Under the conditions prescribed above we now define the coarse-graining time scale $\tilde{t}$ such that
\begin{equation}\label{coarsegrainingtime}
    \tau,\langle t_w \rangle \ll \tilde{t} \ll \tilde{t}_{R},
\end{equation}
where $\tilde{t}_R\equiv {\rm min}(t_R;t_R^\prime)$. The value of $\tilde{t}_R$ is evaluated as minimum between the thermal crossing time and the gravity crossing time of electrons
\begin{equation}\label{relaxationtimes}
        \tilde{t}_R\equiv {\rm min}(t_T;t_{\tilde{g}}) \qquad
        t_T=\frac{2\pi}{\sqrt{T}} \qquad
        t_{\tilde{g}}=\sqrt{\frac{\pi}{\tilde{g}}}
\end{equation}
The temporal coarse-grained phase-space distribution function, defined as the time average of $f_{\alpha}$ over $\tilde{t}$, reads as
\begin{equation}
\begin{gathered}
    \tilde{f}_{\alpha}(\theta,p,\tilde{t})=\langle f_{\alpha} \rangle_{\tilde{t}}=\frac{1}{\tilde{t}}\int_{\tilde{t}} dt f_{\alpha}(\theta,p,t),
\end{gathered}
\end{equation}
and the time-averaged Vlasov dynamics over $\tilde{t}$ becomes
\begin{equation}\label{Vlasovtimeaverage}
    \frac{f_{\alpha}(\theta,p,\tilde{t})-f_{\alpha}(\theta,p,0)}{\tilde{t}}+\frac{p}{M_{\alpha}}\frac{\partial \tilde{f}_{\alpha}}{\partial \theta}+\biggl \langle \tilde{F}_{\alpha}\frac{\partial f_{\alpha}}{\partial p}\biggl \rangle_{\tilde{t}}=0.
\end{equation}
Given the condition $\tilde{t} \ll \tilde{t}_{R}$, during the time interval $\tilde{t}$ the system energy can not be redistributed along the loop. We can therefore approximate $f_{\alpha}$ with its time average $f_{\alpha}=\tilde{f}_{\alpha}$, while the finite difference in the left hand side of \eqref{Vlasovtimeaverage} becomes a time derivative. We then get
\begin{equation}\label{coarsegraineddynamics}
    \begin{gathered}
         \frac{\partial \tilde{f}_{\alpha}}{\partial \tilde{t}}+\frac{p}{M_\alpha}\frac{\partial \tilde{f}_{\alpha}}{\partial \theta}+\tilde{F}_{\alpha}[\tilde{f}_{\alpha}]\frac{\partial \tilde{f}_{\alpha}}{\partial p} =0.
    \end{gathered}
\end{equation}
The thermal boundary at the bottom boundary of the system induces an incoming flux defined at a given time by
\begin{equation}\label{thermalflux}
    J_{T,\alpha}=\int_0^{+\infty}dp\frac{p}{TM_{\alpha}}e^{-\frac{p^2}{2TM_{\alpha}}},
\end{equation}
where the value of $T$ is drawn from the distribution $\gamma(T)$ during the time intervals of length $\tau$; while $T = 1$ during time intervals of length $t_w$ drawn from $\eta(t_w)$.\\
\indent We now compute the temporal coarse-grained flux by taking the time-average of \eqref{thermalflux}, obtaining
\begin{equation}
    J_{\alpha} = \frac{t_{t_w}}{\tilde{t}}\langle J_{T,\alpha} \rangle_{t_{t_w}} + \frac{t_{\tau}}{\tilde{t}}\langle
    J_{1,\alpha} \rangle_{t_{\tau}},
\end{equation}
where $\tilde{t}=t_{t_w}+t_{\tau} \quad t_{t_w}=\sum_{i=1}^{n_p} t_{w_i} \quad t_{\tau}=n_p \tau$ and $n_p$ is the total number of temperature increments within $\tilde{t}$. Assuming $\tau,t_{w_i}\ll\tilde{t}$ implies that the amount of temperature fluctuations in such interval is large enough to justify the use of an ergodic assumption. We can therefore  replace the time average over $t_{t_w}$ with the average with respect to its probability distribution, as
\begin{equation}
    \langle  J_{T,\alpha} \rangle_{t_{t_w}}\\
    =\int_1^{+\infty}dT\gamma(T)J_{T,\alpha}.
\end{equation}
Furthermore, the assumption $\tau,t_w\ll\tilde{t}$ also implies $t_{t_w}=\sum_{i=1}^{n_p} t_{w_i} = n_p \langle t_w \rangle_{\eta}$ where $\langle t_w \rangle_{\eta}$ is given by
\begin{equation}
    \langle t_w \rangle_{\eta} = \int dt_w t_w \eta(t_w)
\end{equation}
so that the coarse-grained flux becomes
\begin{equation}\label{coarsegrainedflux}
    J_{\alpha}=\int_0^{+\infty}dp p \tilde{f}_{\alpha}(p),
\end{equation}
where $\tilde{f}_{\alpha}(p)$ is defined by
\begin{equation}\label{effectivethermostat}
    \tilde{f}_{\alpha}(p)=\frac{A}{M_{\alpha}}\int_1^{+\infty}dT\frac{\gamma(T)}{T}e^{-\frac{p^2}{2TM_{\alpha}}}+\frac{1-A}{M_{\alpha}}e^{-\frac{p^2}{2M_{\alpha}}},
\end{equation}
and $A$ is
\begin{equation}\label{fractionoftime}
    A=\frac{\tau}{\tau +\langle t_w \rangle_{\eta}}. 
\end{equation}
In practice, the dynamics of a two component plasma coupled to a time fluctuating thermal boundary at a given boundary is described, at the temporal coarse-grained level, by a set of Vlasov equations \eqref{coarsegraineddynamics} for the two species coupled to two incoming effective fluxes \eqref{coarsegrainedflux} at the boundary. The latter are generated by the non-thermal distributions $\tilde{f}_\alpha$ given in Eq. \eqref{effectivethermostat}.\\
\indent In the formalism introduced above, it is possible to evaluate an analytic expression for the coarse-grained phase-space distribution functions in the stationary configuration. To do so, one has first to determine such phase-space distributions at the boundary in the stationary configuration. This can be carried out imposing the stationarity and the continuity condition at the boundary, together with the symmetry condition in the momentum space
\begin{equation}
    \frac{\partial \tilde{f}_{\alpha}}{\partial \tilde{t}}=0, \quad
    \mathcal{N}_{\alpha}\tilde{f}_{\alpha}(p)=\tilde{f}_{\alpha}(-\pi,p), \quad \tilde{f}_{\alpha}(-\pi,p)=\tilde{f}_{\alpha}(-\pi,-p), \quad p>0.
\end{equation}
In the expressions above $\mathcal{N}_{\alpha}$ are normalization constants. We can now build up the solution for the entire phase space using the Jeans theorem. We then get 
\begin{equation}\label{VDFmodel}
    \tilde{f}_{\alpha}(\theta,p)=\mathcal{N}_{\alpha}\biggl(A\int_1^{+\infty}dT\frac{\gamma(T)}{T}e^{-\frac{H_{\alpha}}{T}}+(1-A)e^{-H_{\alpha}}\biggl),
\end{equation}
where $H_{\alpha}$ are the mean-field Hamiltonians defined in Eq. \eqref{Mean-field-Hamiltonians}. The constants $\mathcal{N}_{\alpha}$ are fixed by the normalization condition for the two species
\begin{equation}\label{normalization}
    \int_{-\pi}^{\pi}d\theta\int_{-\infty}^{+\infty}dp \tilde{f}_{\alpha}(\theta,p)=1.
\end{equation}
Since $H_{\alpha}$ depends on $\tilde{f}_{\alpha}(\theta,p)$ only through the electrostatic potential $\phi$, this is a self-consistent problem that can be solved with respect to $\phi$ as done for the equilibrium solution \eqref{Thermalesystem} obtaining $\phi=0$. As a consequence, the mean-field-Hamiltonians $H_{\alpha}$ reduce to $\tilde{H}_{\alpha}$ as in Eq. \eqref{Stationarymeanfieldhamiltonians}. The first term in the r.h.s. of Eq.\ \eqref{VDFmodel} is induced by the temperature fluctuations of the thermal boundary and is a superposition of Boltzmann exponential distributions, each of which has a temperature $T>1$ weighted with the probability $\gamma(T)$. Such term introduces the suprathermal contribution and from now on will be referred to as the multitemperature population.\\
 \indent The second term in the r.h.s. of \eqref{VDFmodel} is proportional to a Boltzmann exponential at $T=1$, due to the fact that along each interval of length $t_w$ the thermal boundary is brought back to the temperature $T=1$. From now on we will refer to this term as the thermal population. The relative contribution of the multitemperature and thermal populations is weighted by the factor $A$ given by Eq. \eqref{fractionoftime}, corresponding to the fraction of time in which the thermal boundary is set at one of the temperatures of the multitemperature population.\\
 \indent The 
 shape of the distribution functions depends only on the mean-field Hamiltonians $\tilde{H}_{\alpha}$, since the system is ruled by a Vlasov-type dynamics.\\
\indent Our formalism allows one to discuss the physics in the stationary configuration in terms of velocity filtration. Given the presence of suprathermal tails in the distribution functions (as induced by the multitemperature population) and an external field, the temperature inversion process is expected to take place. This can be easily verified by computing the number densities $n_{\alpha}$ and the kinetic temperatures $T_{\alpha}$ from the equations \eqref{VDFmodel}, using the standard kinetic definitions (see e.g. \citealt{nicholson1983introduction}). For the number densities we get
\begin{equation}\label{NEqdensity}
   \tilde{n}_{\alpha}(\theta)=\frac{A\int_{1}^{+\infty}dT\frac{\gamma(T)}{\sqrt{T}}e^{-\frac{2\tilde{g}}{T}\cos{(\frac{\theta}{2})}}+(1-A)e^{-2\tilde{g}\cos{(\frac{\theta}{2})}}}{A\int_{1}^{+\infty}dT\frac{\gamma(T)}{\sqrt{T}}\int_{-\pi}^{\pi} d\theta e^{-\frac{2\tilde{g}}{T}\cos{(\frac{\theta}{2})}}+(1-A)\int_{-\pi}^{\pi} d\theta e^{-2\tilde{g}\cos{(\frac{\theta}{2})}}},
\end{equation}
while for the kinetic temperatures
\begin{equation}\label{NEqtemperature}
    \tilde{T}_{\alpha}(\theta)=\frac{A\int_{1}^{+\infty}dT\gamma(T)\sqrt{T}e^{-\frac{2\tilde{g}}{T}\cos{(\frac{\theta}{2})}}+(1-A)e^{-2\tilde{g}\cos{(\frac{\theta}{2})}}}{A\int_{1}^{+\infty}dT\frac{\gamma(T)}{\sqrt{T}}e^{-\frac{2\tilde{g}}{T}\cos{(\frac{\theta}{2})}}+(1-A)e^{-2\tilde{g}\cos{(\frac{\theta}{2})}}}.
\end{equation}
In the next Section the above formulas will be evaluated for different choices of model parameters and compared with the output of kinetic numerical simulations.
\section{Comparison with numerical simulations}\label{sec4}
We validated our model with $N-$particle numerical simulations where we integrated the equations of motion \eqref{HMF2SdimensionlessCS} of the system. In the numerical experiments discussed here, as a rule, we fixed $N=2^{21}$ and used a fourth-order symplectic algorithm (see e.g. \citealt{Candy1991}) with fixed time step $\delta t=10^{-4}$. We modelled the incoming energy flux from the fluctuating thermal boundary with the standard technique (see \citealt{ThermalWalls,Landi-Pantellini2001}) such that, when a particle of a given species crosses the boundary, it is re-injected in the system (at the bottom) with a positive velocity sampled from the flux density argument of Eq.\eqref{thermalflux}. We note that, naively re-introducing the particle with a new velocity extracted from a (half) Gaussian at temperature $T$, would break the stationary thermal equilibrium solution, as re-injected particles would have, by construction, a higher chance of having a near zero velocities (see e.g. \cite{LepriMPC2021} and references therein).
\subsection{Stationary state and model parameters}
Motivated by the modelization of the Sun atmosphere, we set the distributions of the temperature fluctuations $\gamma(T)$ and waiting times $\eta(t_w)$ as
 \begin{equation}\label{incrementsandWT}
     \gamma(T)=\frac{1}{T_p}e^{-\frac{(T-1)}{T_p}}, \quad T>1, \quad \eta(t_w)=\frac{1}{\langle t_w \rangle_\eta}e^{-\frac{t_w}{\langle t_w \rangle_\eta}}.
 \end{equation}
The idea is that the chromospheric (i.e. the thermal boundary) temperature fluctuates randomly in time, and the strong temperature increments (i.e., those producing a large $T$), are rather rare (see \citealt{barbieri2023temperature}). This established,  we tune the remaining parameters ($A, \tilde{g},C,T_p$) accordingly. In Figs. \ref{VaryingA}-\ref{Gaussianstatistics} (left panels) we show the time evolution of the  kinetic energies $K_{\alpha}$ (top panels) and of the stratification parameters $q_{\alpha}$ (bottom panels) evaluated in the numerical simulations for different choices of the parameter set as
\begin{equation}\label{Kineticandstrat}
    K_{\alpha}=\frac{1}{N}\sum_{i=1}^N \frac{p_{j,\alpha}^2}{2M_{\alpha}}, \quad q_{\alpha}=\frac{1}{N}\sum_{j=1}^N \cos{(\theta_{j,\alpha})},
\end{equation}
together with their theoretical value (indicated by the straight solid lines) in the stationary configuration, given by
 \begin{equation}\label{KineticandstratinQSS}
     q_{\alpha,SS}=\int_{-\pi}^{\pi}d\theta\int_{-\infty}^{+\infty}dp \cos{\theta}\tilde{f}_{\alpha}(\theta,p), \quad
     K_{\alpha,SS}=\int_{-\pi}^{\pi}d\theta\int_{-\infty}^{+\infty}dp \frac{p^2}{2M_{\alpha}}\tilde{f}_{\alpha}(\theta,p).
 \end{equation}
 In all cases, both $K_{\alpha}$ and $q_{\alpha}$ relax towards their predicted stationary value. In the right panels of the same figures we show the kinetic temperature $T_{e}$ together with the number density $n_{e}$ for the electrons---the profiles for the protons having the same shape---as a function of the spatial coordinate $\theta$. The latter were computed numerically using the standard kinetic definitions (see e.g. \citealt{nicholson1983introduction}) for a single snapshot in the stationary state and then time averaging over subsequent snapshots. In addition, using Eqs. (\ref{NEqdensity}-\ref{NEqtemperature}) we have also computed their theoretical expressions. In all cases presented here, we observe a very good match between the numerical and analytical curves. 
 \subsubsection{The fraction of time $A$}
\begin{figure}
    \centering
    \includegraphics[width=0.99\columnwidth]{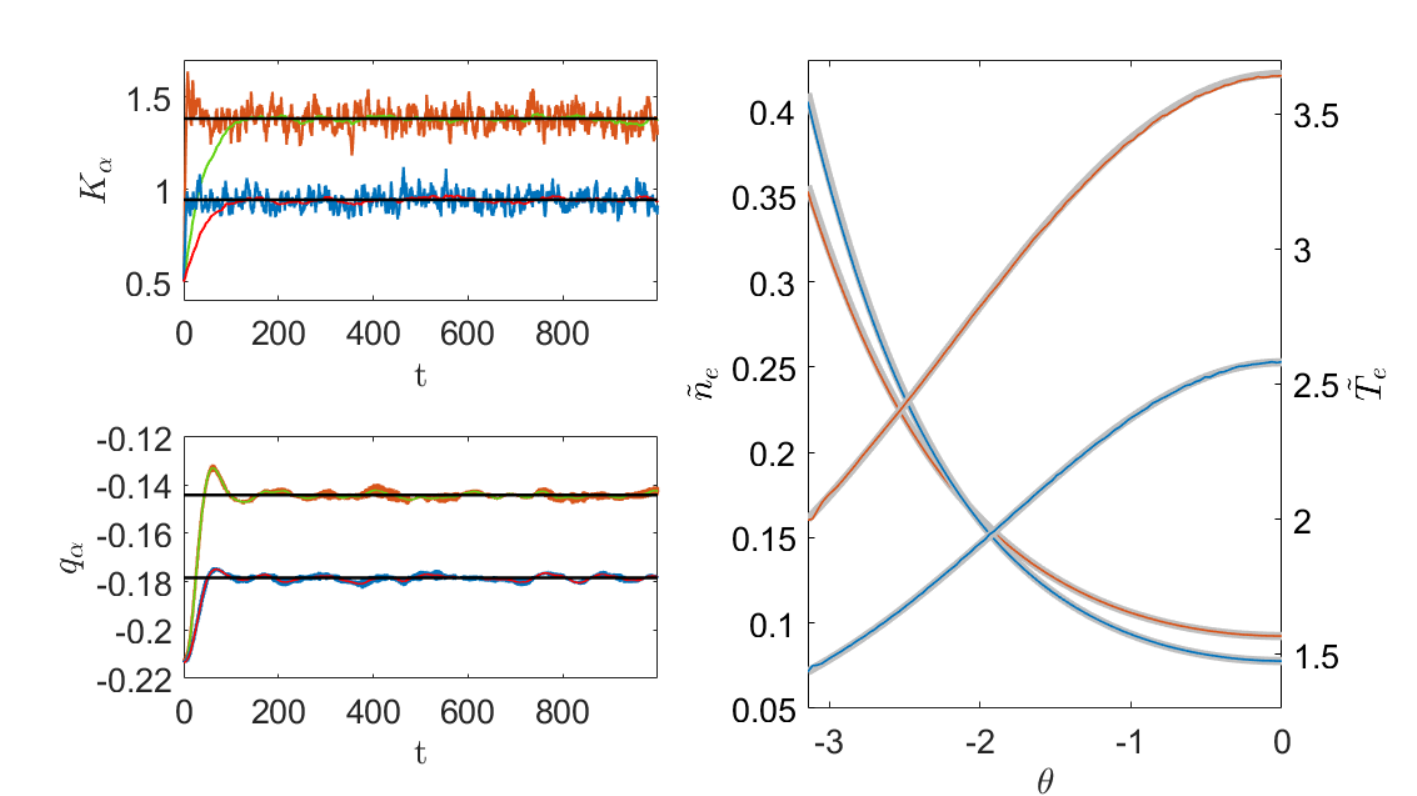}
    \caption{Left panels: Evolution of protons (red for $A=0.25$ and green for $A=0.5$) and electrons (blue for $A=0.25$ and orange for $A=0.5$) kinetic energies $K_\alpha$ (top panel) and stratification parameters $q_\alpha$ (bottom panel) as numerically computed from simulations via Eq.\eqref{Kineticandstrat}, together with theoretical prediction for their mean value (black horizontal lines, see Eq. \eqref{KineticandstratinQSS}). Right panel: numerical density (in blue for $A=0.25$ and in red for $A=0.5$) and temperature (same rule of colours) of electrons vs  curvilinear abscissa of the loop $\theta$. Gray curves denote the corresponding theoretical predictions from analytical formulas Eqs \eqref{NEqdensity}-\eqref{NEqtemperature}.}
    \label{VaryingA}
\end{figure}

 As apparent from Eq. \eqref{fractionoftime}, the parameter $A$ controls the fraction of time in which the thermal boundary is at a given value $T$ sorted from $\gamma(T)$. In Figure \ref{VaryingA} we present two cases:  $A=0.5 (\tau=0.01 ,\langle t_w \rangle_\eta =0.01)$ and $A=0.25(\tau=0.01 ,\langle t_w \rangle_\eta =0.03)$. All the other parameters are fixed to $\tilde{g}=1, T_p=4,C=400,M=100$.\\
 \indent Increasing $A$ corresponds to an increment of the weight of the multitemperature population in Eq. \eqref{VDFmodel} and therefore more energy is injected into the plasma. As a consequence, the plasma becomes less stratified in density (see variation of $\tilde{n}_{\alpha}$ and of the stratification parameters $q_{\alpha}$ in Figure \ref{VaryingA}). For the same reason, the mean value of the kinetic energies $K_{\alpha}$ in the non-equilibrium stationary configuration increases, as the profile of the kinetic temperatures $\tilde{T}_{\alpha}$ does at each in point of the space. 
 \subsubsection{The intensity of temperature increments $T_p$}
\begin{figure}
    \centering
    \includegraphics[width=0.99\columnwidth]{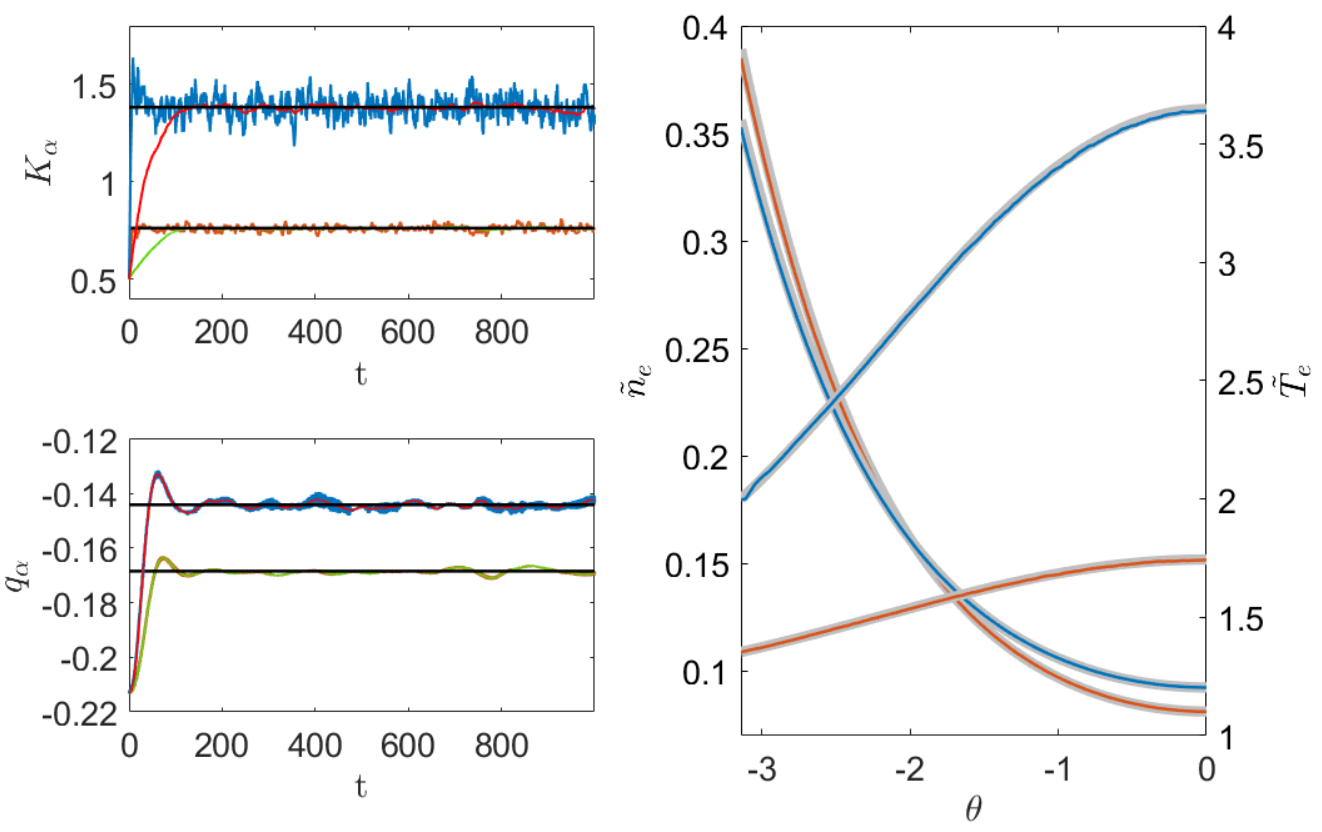}
    \caption{Left panels: Evolution of protons (red for $T_p=4$ and green for $T_p=1$) and electrons (blue for $T_p=4$ and orange for $T_p=1$) kinetic energies $K_\alpha$ (top panel) and stratification parameters $q_\alpha$ (bottom panel) as numerically computed from simulations via Eq. \eqref{Kineticandstrat}, together with theoretical prediction for their mean value (black horizontal lines, see Eq. \eqref{KineticandstratinQSS}). Right panel: numerical density (in blue for $T_p=4$ and in red for $T_p=1$) and temperature (same rule of colours) of electrons vs  curvilinear abscissa of the loop $\theta$. Gray curves denote the corresponding theoretical predictions from analytical formulas Eqs \eqref{NEqdensity}-\eqref{NEqtemperature}.
    }
    \label{VaryingTmean}
\end{figure}

 The parameter $T_p$ (see Eq. \ref{incrementsandWT}) controls the mean value of the (exponential) distribution of the temperature fluctuations. In Figure \ref{VaryingTmean} we present two cases: $T_p=1$ and $T_p=4$. As above, all the other parameters are fixed and are $A=0.5 (\tau=0.01,\langle t_w \rangle_{\eta}=0.01), \tilde{g}=1,C=400,M=100$. Also in this case, increasing the value of $T_p$ makes the energy injected by the multitemperature population to increase. We thus obtain the same behaviour as in the previous case (Figure \ref{VaryingA}) for all quantities. 
\subsubsection{The strength of the electrostatic interactions $C$}
\begin{figure}
    \centering
    \includegraphics[width=0.99\columnwidth]{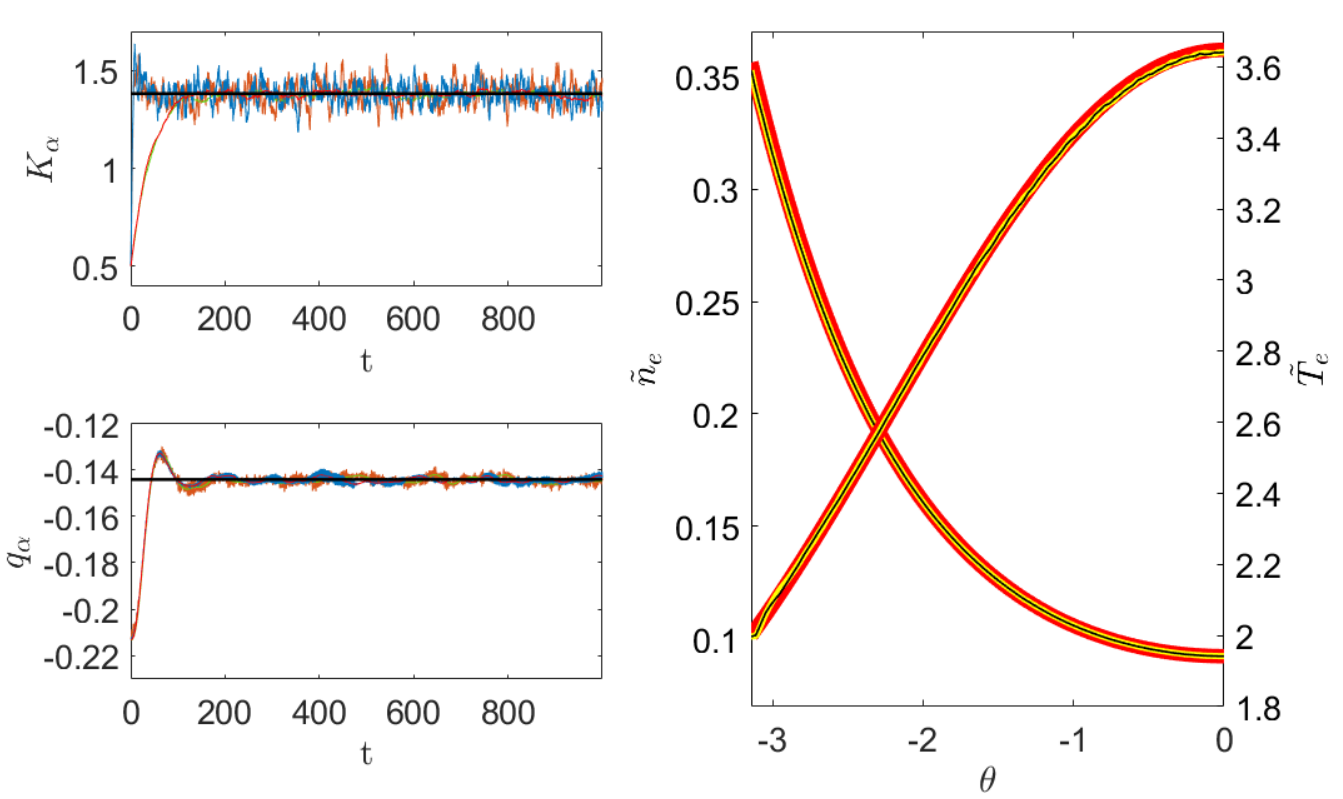}
    \caption{Left panels: Evolution of protons (red for $C=400$ and green for $C=100$) and electrons (blue for $C=400$ and orange for $C=100$) kinetic energies $K_\alpha$ (top panel) and stratification parameters $q_\alpha$ (bottom panel) as numerically computed from simulations via Eq. \eqref{Kineticandstrat}, together with theoretical prediction for their mean value (black horizontal lines, see Eq. \eqref{KineticandstratinQSS}). Right panel: numerical density (in yellow for $C=400$ and in black for $C=100$) and temperature (same rule of colours) of electrons vs curvilinear abscissa of the loop $\theta$. Red curves denote the corresponding theoretical predictions from analytical formulas Eqs \eqref{NEqdensity}-\eqref{NEqtemperature}.
    }
    \label{VaryingC}
\end{figure}

 The parameter $C$ controls the strength of the electrostatic interactions between particles.
 In Figure \ref{VaryingC} we report the two cases $C=100$ and $C=400$. All the other parameters are fixed such that $A=0.5 (\tau=0.01,\langle t_w \rangle_{\eta}=0.01), T_p=4,\tilde{g}=1,M=100$. This Figure clearly shows that the electrostatic interaction $C$ does not play any significant role in shaping the inverted density-temperature profiles, as obvious from the theory since $\phi=0$, cfr. Sect. \ref{sec3}.
 \subsubsection{The stratification parameter $\tilde{g}$}
\begin{figure}
    \centering
    \includegraphics[width=0.99\columnwidth]{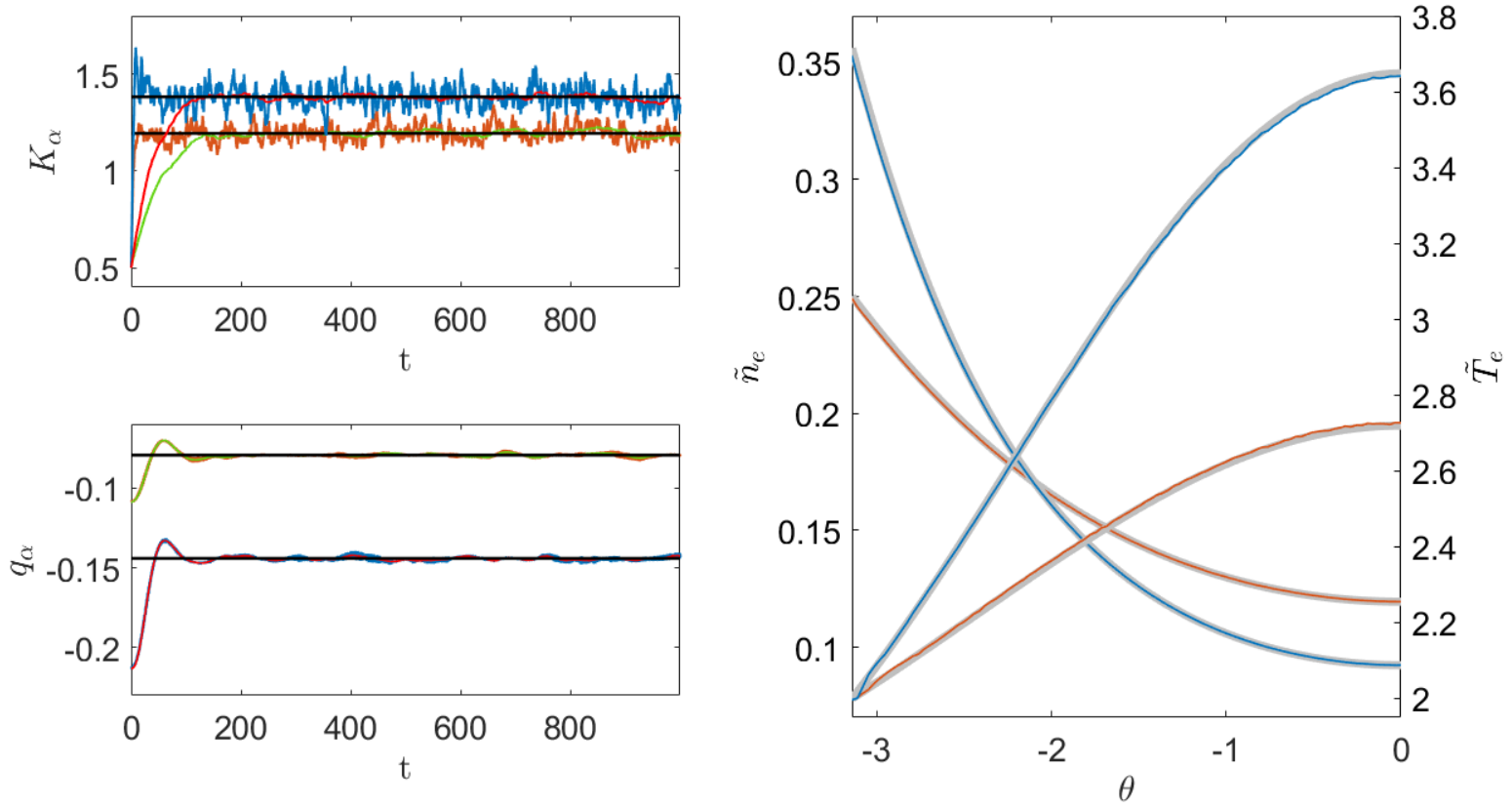}
    \caption{Left panels: Evolution of protons (red for $\tilde{g}=1$ and green for $\tilde{g}=0.5$) and electrons (blue for $\tilde{g}=1$ and orange for $\tilde{g}=0.5$) kinetic energies $K_\alpha$ (top panel) and stratification parameters $q_\alpha$ (bottom panel) as numerically computed from simulations via Eq. \eqref{Kineticandstrat}, together with theoretical prediction for their mean value (black horizontal lines, see Eq. \eqref{KineticandstratinQSS}). Right panel: numerical density (in blue for $\tilde{g}=1$ and in red for $\tilde{g}=0.5$) and temperature (same rule of colours) of electrons vs curvilinear abscissa of the loop $\theta$. Gray curves denote the corresponding theoretical predictions from analytical formulas Eqs \eqref{NEqdensity}-\eqref{NEqtemperature}.
    }
    \label{Varyingg}
\end{figure}

 This parameter determines the strength of the external stratification field in units of the thermal energy $k_B T_0$ of the thermal boundary (i.e., the chromosphere). In Figure \ref{Varyingg} we present two cases:  $\tilde{g}=0.5$ and $\tilde{g}=1$. All the other parameters are as before.
 The expressions of the distribution functions of the theory reported in Eq. \eqref{VDFmodel} depend only on $\tilde{H}_\alpha$. At the bottom (i.e., $\theta=-\pi$), the mean-field Hamiltonians $\tilde{H}_{\alpha}$ are the same for both species and read
 \begin{equation}
     \tilde{H}_{\alpha}(p,\theta=-\pi)=\frac{p^2}{2M_{\alpha}}.
 \end{equation}
 Therefore, the two distribution functions have a fixed width in momentum space that is independent of the specific value of $\tilde{g}$. Hence, also the kinetic temperature at $\theta=-\pi$ is independent of $\tilde{g}$, as apparent in Figure \ref{Varyingg}.\\ 
\indent In the right plot of Figure \ref{Varyingg} we show that, by increasing the value of $\tilde{g}$, the kinetic temperature grows with a stronger gradient going from the bottom ($\theta=-\pi$) to the top ($\theta=0$). In practice, as implied by Eq. \eqref{VDFmodel}, with higher values of $\tilde{g}$, only the Boltzmann factors with large values of the temperature in the multitemperature population can give a contribution at higher altitudes. In practice, from all the possible value extracted from $\gamma(T)$, only high $T$ temperature fluctuations give a sufficient amount of kinetic energy to the particles so that they can climb the gravitational well up to the top of the system $\theta=0$. We note that, this is essentially Scudder's gravitational filtering mechanism.
 \subsubsection{Varying the temperature fluctuations distribution $\gamma(T)$}
\begin{figure}
    \centering
    \includegraphics[width=0.99\columnwidth]{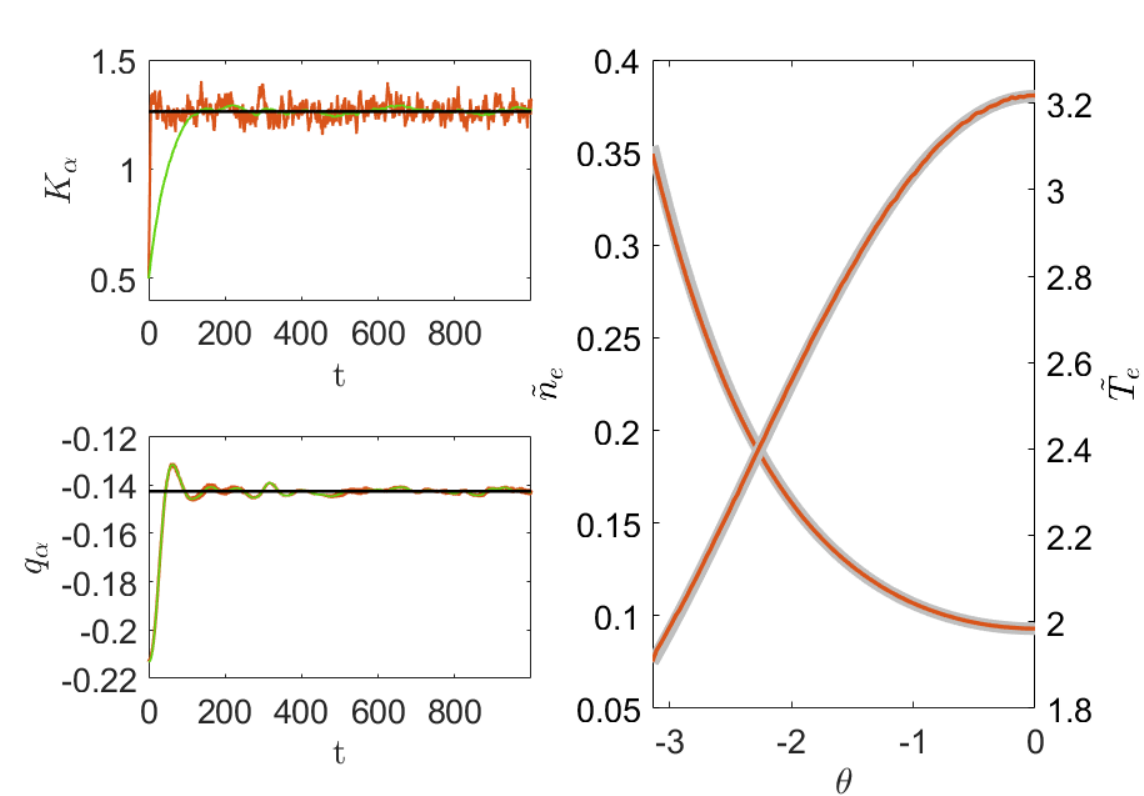}
    \caption{Left panels: Evolution of protons (green) and electrons (red) kinetic energies $K_\alpha$ (top panel) and stratification parameters $q_\alpha$ (bottom panel) as numerically computed from simulations via \eqref{Kineticandstrat}, together with theoretical prediction for their mean value (black horizontal lines, see Eq. \eqref{KineticandstratinQSS}). Right panel: numerical density and temperature (red) of electrons vs curvilinear abscissa of the loop $\theta$. Gray curves denote the corresponding theoretical predictions from analytical formulas Eqs \eqref{NEqdensity}-\eqref{NEqtemperature}.
    }
    \label{Gaussianstatistics}
\end{figure}

 Finally, we change the distribution of temperature increments $\gamma(T)$. We consider here the case of a one sided Gaussian $\gamma(T)$ distribution
\begin{equation}
    \gamma(T)=\frac{2}{\sqrt{2\pi T_p^2}}e^{-\frac{(T-1)^2}{2T_p^2}} ,\quad T>1.
\end{equation}
In Figure \ref{Gaussianstatistics} we show the case for the following values of the system's  parameters: $A=0.5 (\tau=0.01,\langle t_w \rangle_\eta =0.01),\tilde{g}=1,T_p=4,C=400,M=100$. As expected, temperature increments still give origin to temperature inversion independently of the specific shape of $\gamma(T)$.
\subsection{Limits of the theoretical approach}
\begin{figure}
    \centering
    \includegraphics[width=0.99\columnwidth]{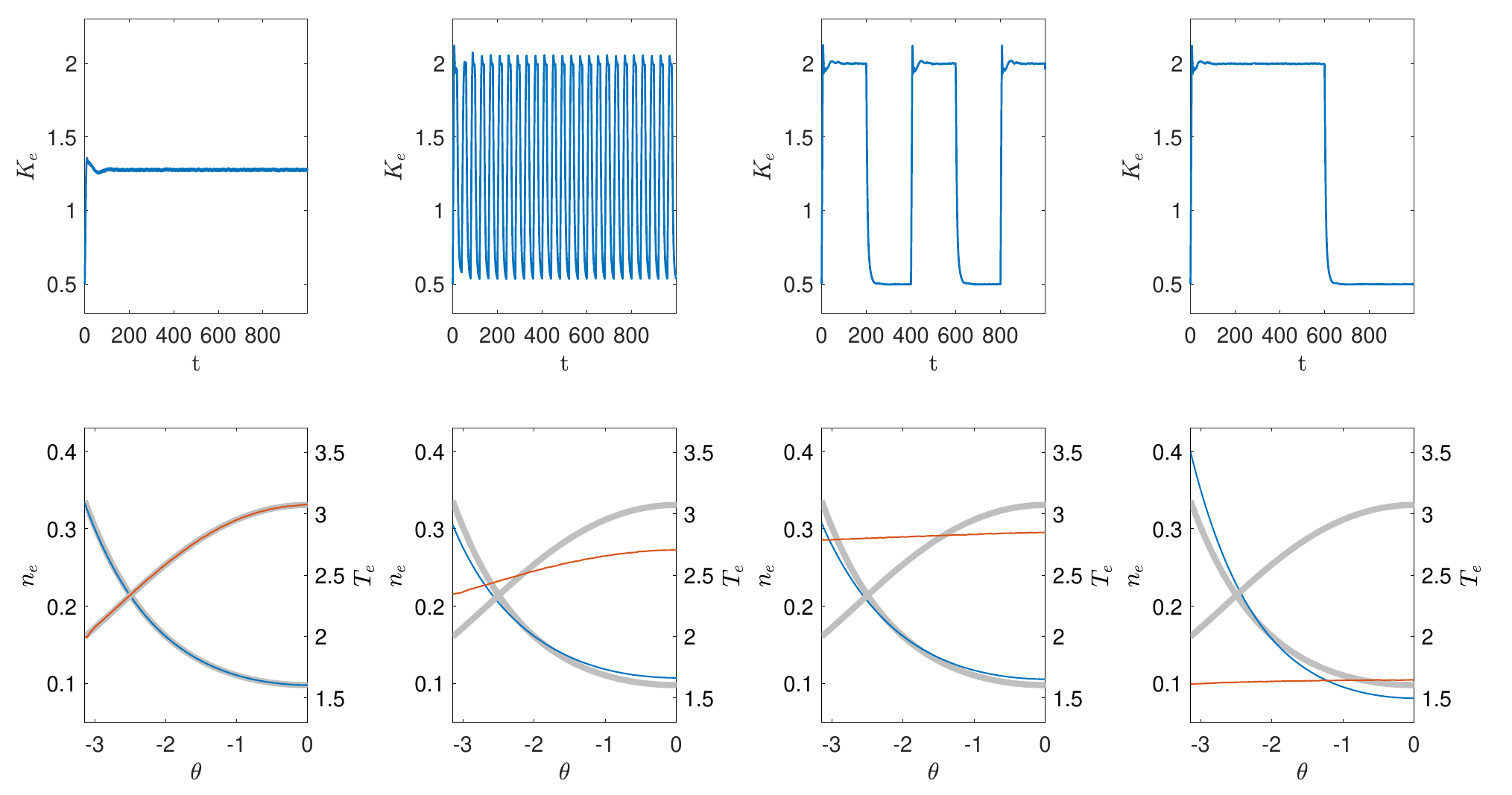}
    \caption{First row: kinetic energies of the electrons for four different couples of values of $\tau=t_w={0.1,20,200,600}$ from left to right. Second row: electrons temperature-density profiles numerically evaluated time averaging over the interval $t_1=300$ and $t_2=1000$ (red for temperature and blue for density) together with the theoretical predictions in grey and calculated via Eqs. \eqref{NEqdensity}-\eqref{NEqtemperature}.}
    \label{Limitstheory.pdf}
\end{figure}

\begin{figure}
    \centering
    \includegraphics[width=0.99\columnwidth]{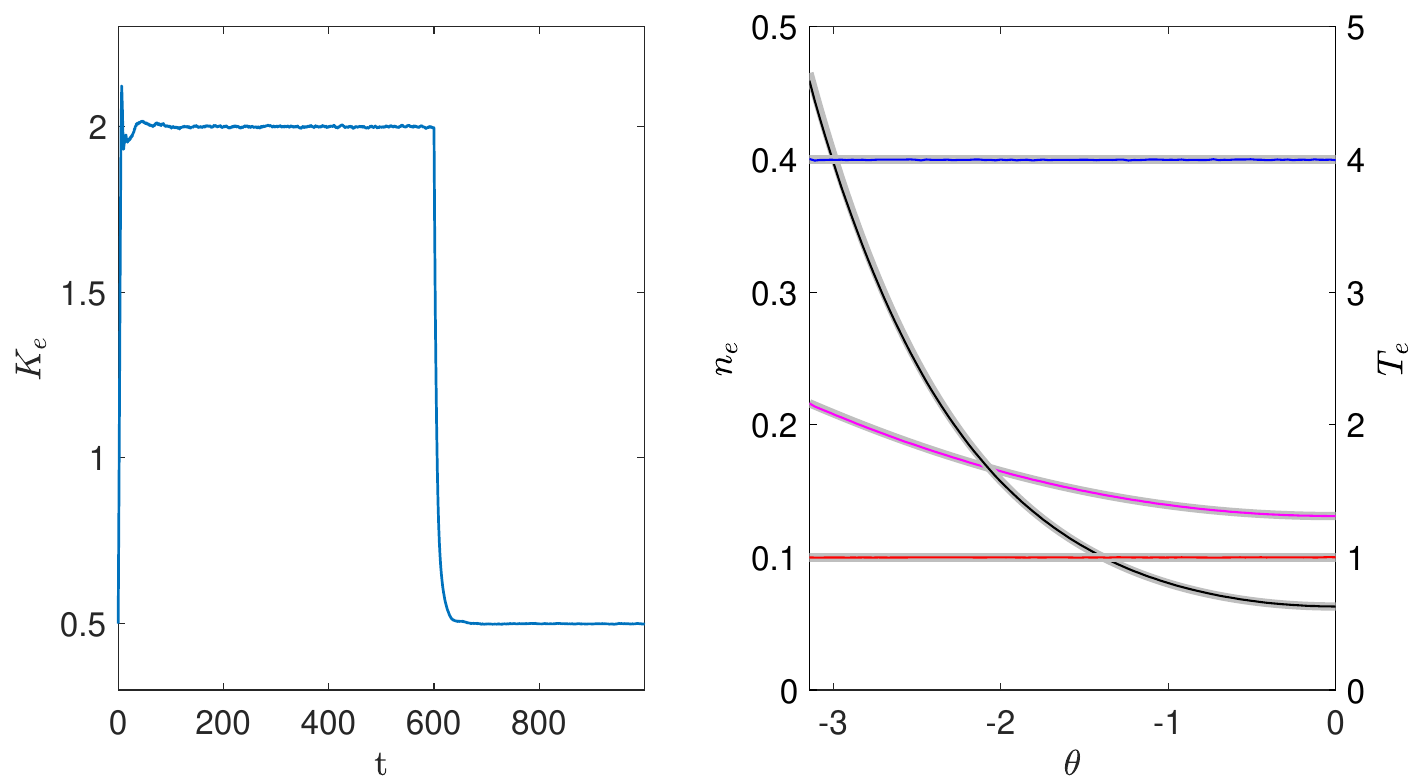}
    \caption{Left: electrons Kinetic energy as a function of time. Right: electrons densities and temperatures for the two thermal configurations $T=T_p=4$ and $T=1$. In blue and magenta respectively the numerical temperature and the numerical density for the case $T=T_p=4$, together with their respective theoretical equilibrium counterparts in grey and calculated via Eqs. \eqref{Eqdensity}-\eqref{Eqtemperature}. In red and black respectively the numerical temperature and the numerical density for the case $T=1$, together with their respective theoretical equilibrium counterparts in grey and calculated via Eqs. \eqref{Eqdensity}-\eqref{Eqtemperature}.}
    \label{Eqdensitytemperatures.pdf}
\end{figure}

We now discuss the limits of the theoretical approach, using numerical simulations. For the sake of simplicity, we focus on the case in which the temperature oscillates between the the two values $T=1$ and $T=T_p$ with a fixed waiting time $t_w$. In this setup the distributions $\gamma(T)$ and $\eta(t_w)$ are given by
\begin{equation}
    \gamma(T)=\delta(T-T_p), \qquad \eta(t_w)=\delta(t-t_w).
\end{equation}
In Figure \ref{Limitstheory.pdf}, we report the results obtained by varying 
$\tau$, with $\tau=t_w$. The other parameters are fixed to $C=400,T_p=4,\tilde{g}=1,M=100$.\\ 
\indent  We observe how increasing both $\tau$ and $t_w$ and retaining the ratio $A=\frac{\tau}{\tau + t_w}$ fixed to $0.5$, the theoretical equations computed using Eq. \eqref{VDFmodel} start failing in reproducing the density-temperature profiles evaluated from the numerical simulations. In particular, in the left panels the time scales $\tau$ and $t_w$ are so small that the particles do not have enough time to relax to a thermal configuration, neither at $T=T_p$ or at $T=1$. The system is therefore locked in a non-equilibrium stationary configuration and the inverted density-temperature profiles computed numerically are well reproduced by the theoretical counterparts. Technically speaking, the inverted density-temperature profiles shown in Fig. \ref{Limitstheory.pdf} are obtained by time-averaging over several snapshots from the $t_1=300$ to the end time $t_2=1000$. Since the system is in a stationary state these profiles are independent on the length and on the location along the stationary state of the interval $t_2-t_1$.\\ 
\indent For the other cases, the density-temperature profiles are numerically evaluated in the same way. As both $\tau$ and $t_w$ increase, the kinetic energy starts to develop large amplitude oscillations. This arises from the fact that when moving towards a regime (shown in the two right columns of the Figure \ref{Limitstheory.pdf}) in which both $\tau$ and $t_w$ are large enough (i.e. comparable to the relaxation time) such that the system evolves in time passing from a thermal configuration at $T=T_p$ to a thermal configuration at $T=1$ (see Figure \ref{Eqdensitytemperatures.pdf}).\\ 
\indent Of course, in all cases except when $\tau=t_w=0.1$, since the system is no longer locked in a stationary configuration, the theory can not be applied and thus the numerically recovered density-temperature profiles evaluated via time averaging do not match the theoretical counterparts. Moreover, since the system is no more locked in a stationary state, the numerical density-temperature profiles depend explicitly on the length and on the location of the time average interval $t_2-t_1$ for the specific run.
\begin{figure}
    \centering
    \includegraphics[width=0.99\columnwidth]{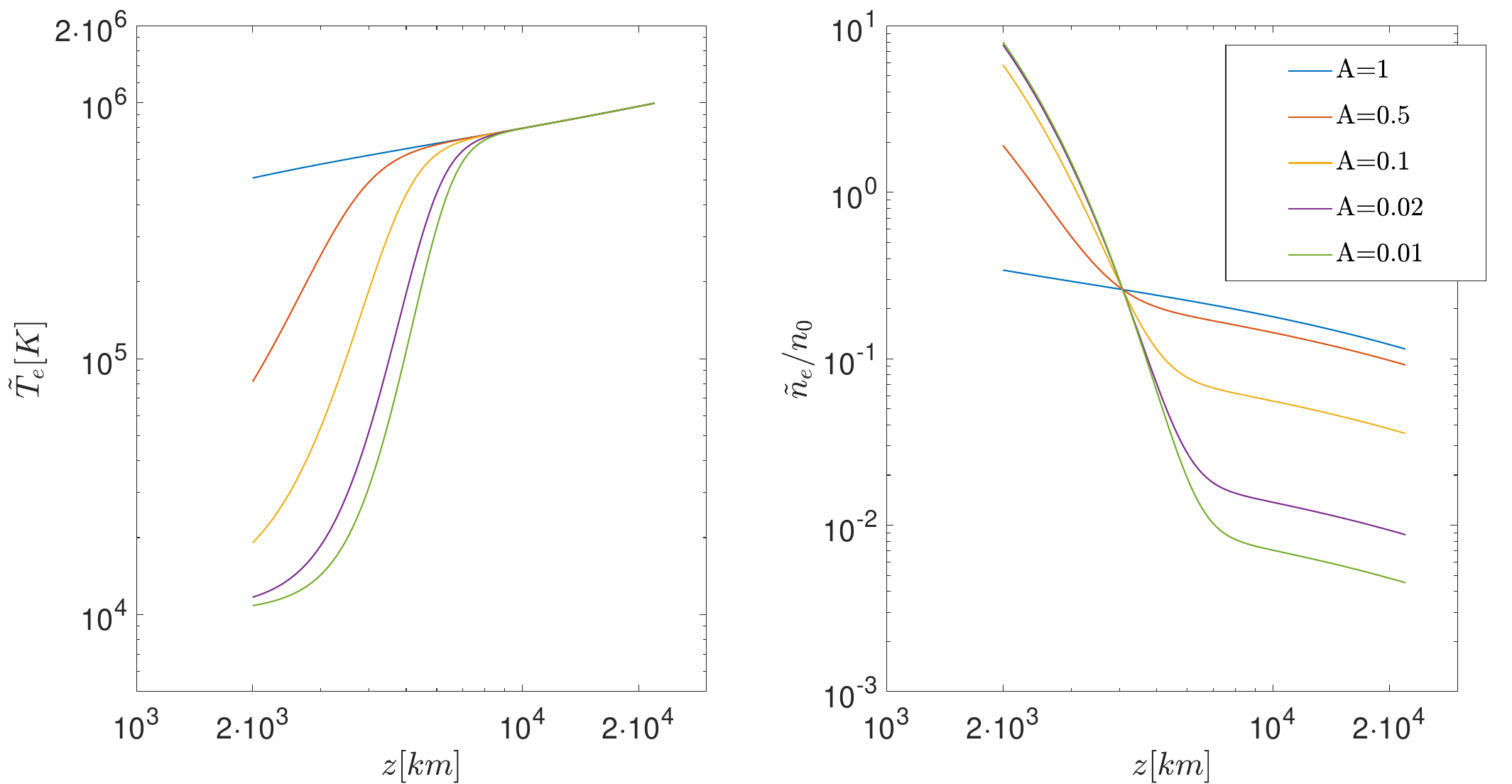}
    \caption{Kinetic temperature (in K, left panel) computed via \eqref{NEqtemperature} and number density computed via \eqref{NEqdensity} (right panel) scaled by the mean number density $n_0=2.5\cdot 10^9 \mathrm{cm}^{-3}$ as a function of the height (in km) for different values of $A$ passing from $A=1$ to $A=0.01$ as listed in the legend.}
    \label{TvsnModel2}
\end{figure}
\begin{figure}
    \centering
    \includegraphics[width=0.99\columnwidth]{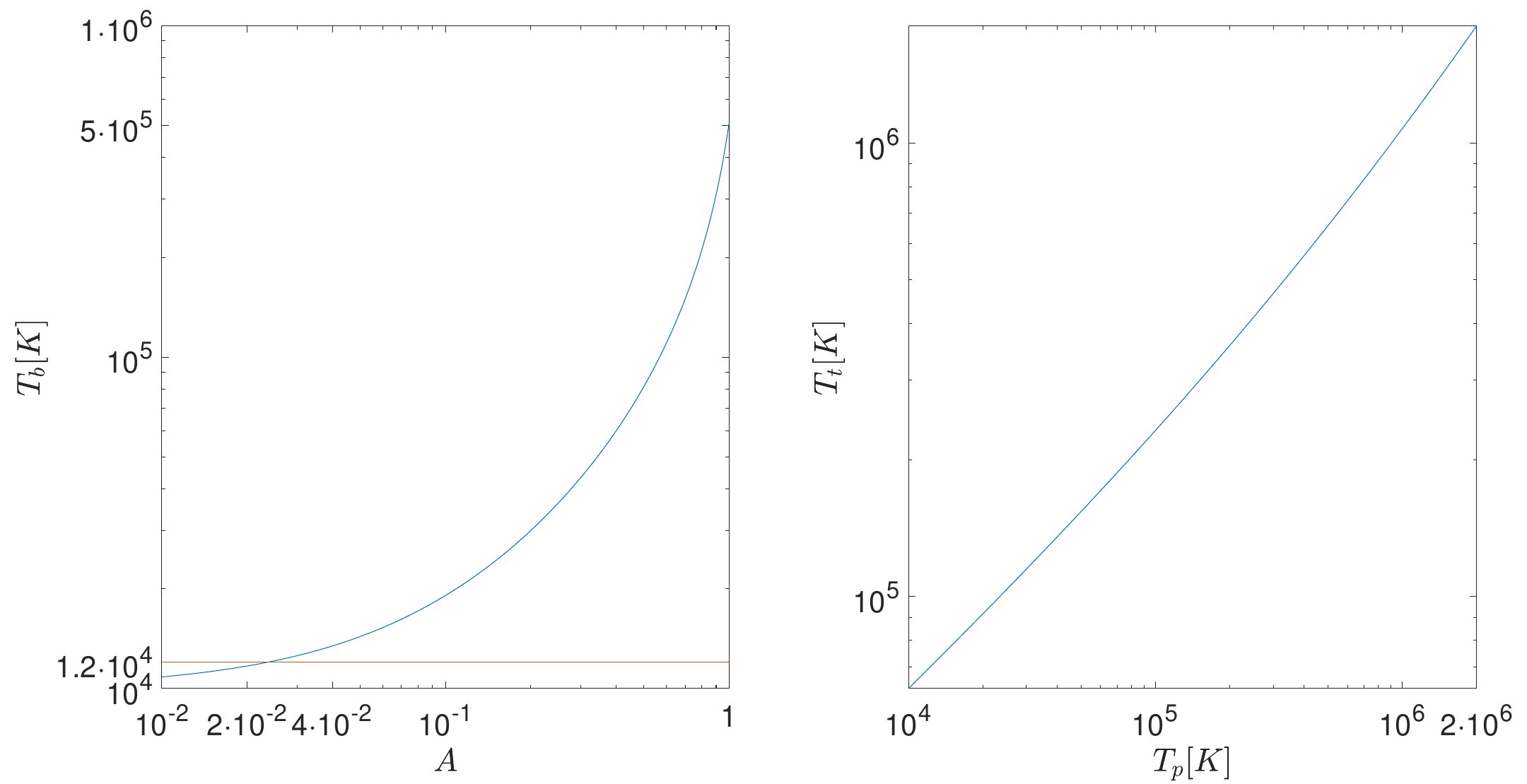}
    \caption{Left panel: temperature at base $T_b$ of the atmosphere $z=0$ computed via Eq. \eqref{NEqtemperature} against the parameter $A$ Eq. \eqref{fractionoftime}. The red horizontal line is $T_{b,obs}=1.2\times 10^4$ K. Right panel: the temperature at the top $T_t$ computed via Eq. \eqref{NEqtemperature} as function of the mean-value of the temperature fluctuations $T_p$ for $A=1$. }
    \label{TbTt2}
\end{figure}
\begin{figure}
    \centering
    \includegraphics[width=0.99\columnwidth]{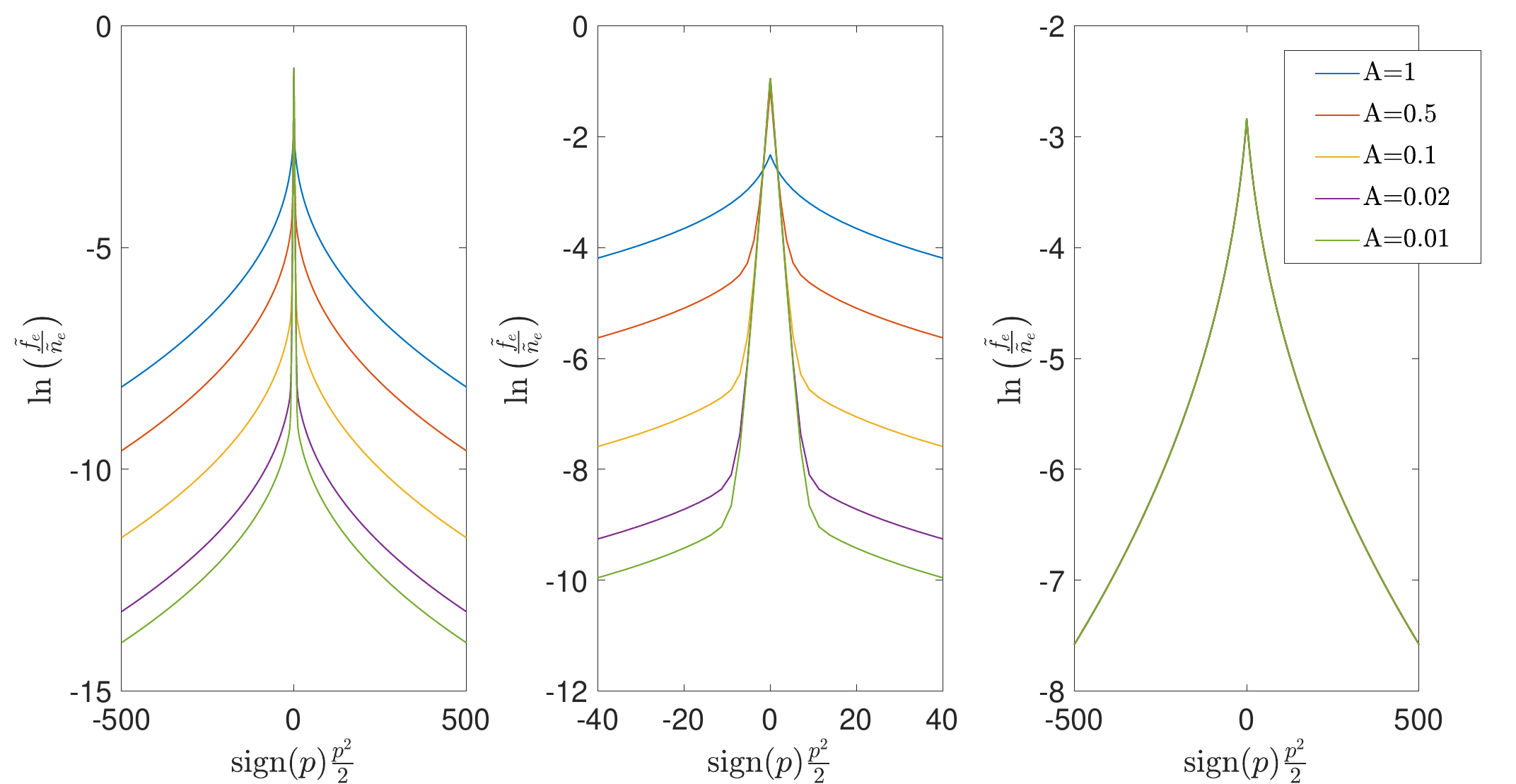}
    \caption{Left panel: electron velocity distribution functions (VDFs) computed via \eqref{VDFmodel} normalized by the electron number densities (computed via \eqref{NEqdensity}) at the base of the atmosphere at $z=z_1=2.3\times 10^3 \mathrm{km}$ (the base of the transition region) for different values of $A$. In the central panel a magnification of the central region of the same VDFs are shown to emphasise the central thermal (Gaussian) core. In the right panel are depicted the electron VDFs in the corona $z=z_3$ for the same values of $A$ of the other panels. In this panel all the electron VDFs are collapsed onto each other because of velocity filtration.}
    \label{VDFvariousA}
\end{figure}
\begin{figure}
    \centering
    \includegraphics[width=0.99\columnwidth]{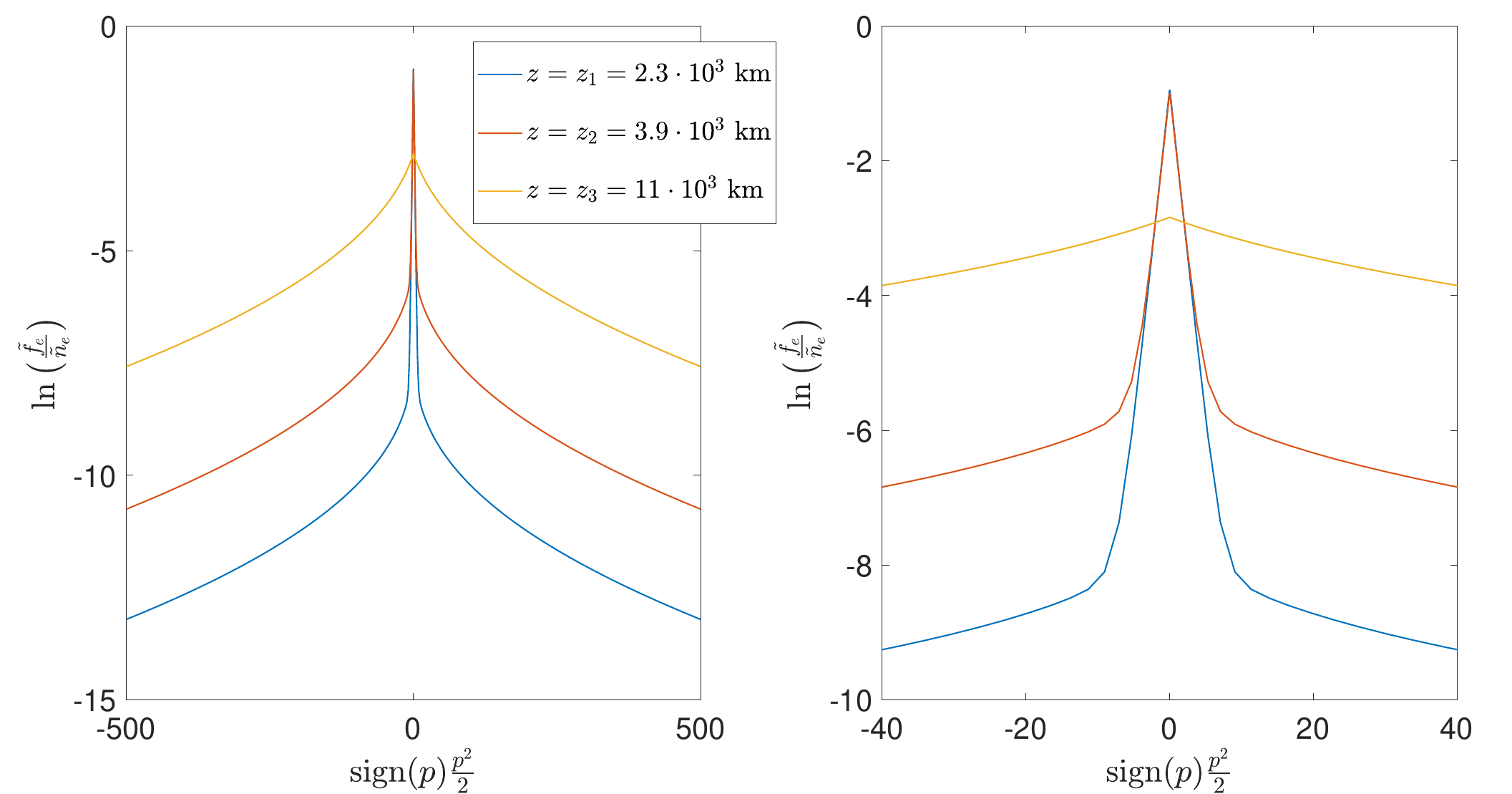}
    \caption{Left panel: electron DFs (computed via \ref{VDFmodel} normalized by the electron number densities (computed via \ref{NEqdensity}) for $A=0.02$, computed at three increasing heights ($z=z_1,z_2,z_3$) listed in the legend as a function of the signed kinetic energy. In the right panel, a magnification of the central region of the same DFs is shown to appreciate the disappearance of the Gaussian profile with height, as expected when velocity filtration is at work.}
    \label{VDFvariousz}
\end{figure}
\section{Application to the solar atmosphere}\label{sec5}
Here we show the application of the temporal coarse-graining to predict the inverted density-temperature profiles of the Sun atmosphere. To compute such profiles we have used Eqs.\ \eqref{NEqdensity}-\eqref{NEqtemperature}, and Eqs.\ \eqref{VDFmodel} for the distribution functions. We put the base of the model at $z_b=2\times 10^3$ km (i.e., at the base of the transition region in the Sun atmosphere) and the top at $z_t=2.2\times 10^4$ km (i.e., in the corona). We then fix the length $2L$ of half of the loop via the following equation
\begin{equation}
    2L=(z_t-z_b)\pi=2\pi \times 10^4 {\rm km}, \quad z=\frac{2L}{\pi}\cos{\biggl(\frac{\pi x}{2L}\biggl)},
\end{equation}
that corresponds to the typical length of a coronal loop in the solar atmosphere. With such choice, the dimensionless parameter $\tilde{g}$ is fixed to $\tilde{g}=16.64$.  For the distribution of the temperature fluctuations and the waiting times, we use Equations \eqref{incrementsandWT} discussed in section 4 above. 
Thus, we are now left with the two free parameters $T_p$ and $A$. We initially fix the value of $A$ to $1$ and vary $T_p$.\\
\indent As shown in Section 4, increasing the value of $T_p$ results in increasing the values of the temperature profile. In order to have a temperature of the order of $10^6$ K at the top of the loop, we must increase $T_p$ up to a value around $10^6$ K (as shown in the right panel of figure \ref{TbTt2}). Therefore, from now on, we fix $T_p=90$ (that correspond to a temperature of $9\times 10^5$ K). From the left panels of figures \ref{TbTt2} and \ref{TvsnModel2} we observe that for $A=1$ we have a temperature at the base of the order of $T_b=5\times 10^5$ K. Such value is far greater than the observed one, the latter being smaller than $T_{b,obs}=1.2\times 10^4 {\rm K}$, for the chromospheric temperature varies between $8\times 10^3$ and $1.2\times 10^4$ K, see \cite{Molnar_2019}.\\
\indent In the previous section we have shown how the temperature at the base of the loop $T_{b}$ decreases with decreasing $A$ (see figure \ref{VaryingA}). In Figure \ref{TbTt2} we plot $T_b$ as a function of $A$. We observe that at around A=$0.022$ the temperature $T_b$ crosses the upper limit value $T_{b,obs}$.
Combining these two properties that relate $T_p$ to $A$, in  Figure \ref{TvsnModel2} we show how the inverted density-temperature profiles with $A=0.02$ and $0.01$ (i.e. very rare temperature increments with $T_p = 90$) are similar to the one observed in the Sun atmosphere. These profiles start from $T_b \sim 1.1\times 10^4$ K, they have an initial rapid rise in temperature (the transition region) followed by a slower increase in the above region (i.e. the solar corona).\\
\indent This change of the shape caused by varying the parameter $A$ can be understood in terms of velocity filtration, as it is apparent from the structure of the electron velocity distribution functions (VDFs) at the base of the model ($z=z_1=2.3\times 10^3$ km) as shown in Figure \ref{VDFvariousA} where we plot in a semilogarithmic scale the distribution of the signed electron kinetic energy $\tilde{f}_e (\text{sign}(p)\,p^2/2)$ divided by the density. By doing this, it becomes clearer whether the distribution function in question is thermal or not, since for a thermal distribution (i.e. a Gaussian) one would get two straight lines symmetric with respect to zero. As can be seen from the left and the middle panels of Figure \ref{VDFvariousA}, the thermal (Gaussian) part of the electron VDFs increasingly dominates the core of the distribution, as the value of $A$ decreases. This is the reason why the temperature at the base decreases crossing $T_{b,obs}=1.2\times 10^4$ K and tends to $T_0=10^4$ K. Moreover, by fixing $A=0.02$, from Figure \ref{VDFvariousz} we observe that, in virtue of the energy conservation (thanks to the velocity filtration mechanism), the cold central thermal core of the electron DFs at the base of the model $z=z_1=2.3\times 10^3$ km progressively disappears by passing through the transition region $z=z_2=2.3\times 10^3$ km and is totally filtered out in the corona $z=z_3=1.1\times 10^4$ km. As a consequence of this, one has a rapid increase of temperature through the transition region and a slow increase of temperature in the corona.\\
\indent In right panel of Figure \ref{VDFvariousA} we can observe that all the electron DFs rescaled by the density in the corona at $z=z_3=1.1\times 10^4$ km for all the values of $A$ collapse onto each other, due to the fact that the central thermal core has been filtered out in the corona itself. Since the rescaled distribution functions are the same for all values of $A$, then the temperatures collapse to the same curve in the corona as shown in the left panel of Figure \ref{TvsnModel2}.\\
\indent The number density profiles are inverted with respect to the temperature profiles with an initial strong negative gradient (the transition region) followed by a slower decrease within the corona. By decreasing the value of $A$, the number density at the bottom of the system increases and at the same time the number density at the top decreases. Again this is due to the fact that the cold population in Equation \eqref{VDFmodel} becomes more and more pronounced for decreasing values of $A$, that is, we now have more particles that are not able to climb higher in the gravity well. Since the total number of particles is fixed independently on the values of $A$ as dictated by the normalization condition \eqref{normalization} then, if the density at the bottom increases, the density at the top (the corona) should naturally decrease.
\subsection{Observational evidences of temperature increments and physical limits of our modelling}

We now discuss the limits of our modelling approach. In our model, we only considered one-sided temperature fluctuations, based on the assumption that a physical process heats up the chromosphere for a finite amount of time. 
Such assumption justifies the use of positive temperature increments only. Observational evidences of physical processes that can produce such increments will be presented later in this section. We stress the fact that, even if we decrease the temperature below the background value, we still have a temperature inversion. In this case, the ``hot" particles are the ones produced by the background temperature and the ``cold" ones are produced by the temperature decrements. Therefore, velocity filtration is still active, and the temperature starts from a value that is smaller compared to the background due to the colder particles but raises to the background value.

In conclusion, the temperature fluctuations drive the above collisionless plasma environments towards a non-equilibrium stationary configuration with temperature inversion that can be described by our temporal coarse-graining method.

This is true only if the time scales of the temperature fluctuations are much smaller than the relaxation time $t_R$ (computed via Eq. \eqref{relaxationtimes}, roughly $10$~s for the parameters that we have considered above). 

Moreover, to reach  coronal temperatures of $10^6$~K and keep at the same time the temperature at the base of the transition region around $10^4$~K, the fraction of time in which the chromospheric temperature increments must be as large as one-million degree must be around $A \sim 0.02$. More precisely, with this value of $A$, the temperature at the base is around $T_b \sim 1.2 \times 10^4$ this is compatible with ALMA observations of temperatures $1.2 \times 10^4$~K \citep{Molnar_2019}. 

Indeed, short-lived and intense brightenings are observed on the Sun surface \citep{Dere:1989ux,Teriaca:2004wy,Peter:2014uz,Tiwari:2019us,Berghmans:2021wl}.
Among all of these the so-called campfires, observed in extreme UV imaging, have temperatures of $\approx 10^6$ K. Other explosive events appearing in $\mathrm{H}\alpha$ line widths have smaller temperatures, around $2 \times 10^5$ K, but are ten times more frequent \cite{Teriaca:2004wy}. This trend justify the use of a decreasing exponential distribution for the temperature increments. Anyhow, current measurements have a time resolution of the order of few seconds (i.e. of the order of the relaxation time $t_R$) so that even if the intensity of such increments could be compatible with recently observed explosive events, they remain unresolved because of their rapid time duration. 
Interestingly, recent EUV solar observations \citep{Raouafi_2023} show that a possible physical mechanism that is at work in low atmosphere is magnetic reconnection. In particular they show how jetting activity at the base of the solar corona driven by discrete small-scale magnetic reconnection events (i.e. continuous sources of temperature increments) can produce a flow of matter that propagates from the corona up to the heliosphere so that they could be at the origin of the coronal heating and of the solar wind acceleration.

The most questionable aspect of our treatment is that the chromosphere is modelled as fully collisional and cuncurrently the overlying plasma (the transition region and the corona) is modelled as collisionless. Due to the low values of the Knusden number $K_n$ ($K_n\sim 10^{-2}-10^{-3}$ in the transition region and $K_n\sim 10^{-1}$ in the corona) the collisions are certainly present in both the transition region and the corona. 
Clearly, a fully self-consistent treatment of Coulomb collisions in the whole solar atmosphere is beyond the scope of this work. Anyhow, it is worth to briefly address this problem. Thanks to the $1/r$ (where $r$ is the inter-particle distance) nature of electrostatic potential, the electron-electron mean free path in a plasma strongly depends on the particles velocity $v$, more specifically it is proportional to $v^4$. In our procedure we have "cold" particles generated by the background temperature $T_0$ and "hot" particles generated by the temperature increments. We expect that only "cold" particles would be strongly affected by collisions so that the VDFs would be closer to thermal at low energies, while non-thermal features associated with the "hot" particles, which are the ones selected by gravitational filtering, would be able to reach the corona (\cite{Landi-Pantellini2001}). Moreover, velocity filtration should become more efficient in presence of Coulomb collisions, thus producing a transition region even steeper than the one produced by our model and closer to the observed one.

\section{Summary and perspectives}\label{sec6}
\indent In this work, we have addressed the question of generating suprathermal distribution functions and temperature inversion in a confined plasma structure by means of temperature increments at its edges produced by some heating events. We tackled this problem introducing a novel formalism to treat the plasma dynamics in contact with a thermal boundary whose temperature fluctuates in time recently studied by \citealt{barbieri2023temperature}. We have shown how the multi-component (in the case discussed here, electrons and protons) plasma dynamics can be efficiently modelled in terms of the temporal-coarse-grained distribution functions $\tilde{f}_{\alpha}$ for the two species. The dynamics of the latter is governed by the effective system of Vlasov equations \eqref{coarsegraineddynamics} in contact with the non-thermal distributions $\tilde{f}_{\alpha}(p)$ at its boundary \eqref{effectivethermostat}.\\
\indent We have obtained analytical expressions for the distribution functions of the two species in the non-equilibrium stationary configuration \eqref{VDFmodel}, in terms of the mean field Hamiltonians $\tilde{H}_{\alpha}$ Eq.\eqref{Stationarymeanfieldhamiltonians}. In this setup we can interpret the anti-correlated density-temperature profiles in terms of velocity filtration, in analogy to the Scudder mechanism. However, here the suprathermal tails are not imposed as in the Scudder
approach but are self-consistently produced by the temporal variations of the temperature of the thermal boundary (the chromosphere) yielding particles with different temperatures that mix together giving rise to the multitemperature contribution to the coarse-grained distribution functions in
Eqs.\ \eqref{VDFmodel}.\\
\indent We have 
tested the theoretical predictions for temperature inversion against numerical simulations, finding an excellent agreement between the theory and the numerical results. In addition, we have applied our theoretical formalism to the specific case of the Sun atmosphere. We have shown how decreasing the percentage of time of duration of the temperature increments (i.e. decreasing $A$) a transition region  naturally forms in the inverted density-temperature profile (see Figure \ref{TvsnModel2}).  Moreover, in the condition of very rare temperature fluctuations $A\sim 0.02$ at $T_p \sim 10^6$ K  we were able to produce an inverted temperature-density profile very similar to the one observed in the Sun atmosphere with a base temperature below $1.2\times 10^4$ K (as observed e.g. by \citealt{Molnar_2019}), transition region (wider than the observed one) and then a million-kelvin corona.\\
\indent In the present work we have derived the coarse grained Vlasov dynamics \eqref{coarsegraineddynamics} coupled to the effective incoming flux \eqref{coarsegrainedflux} in the one-dimensional case and using the HMF modelization. The extension of our procedure to systems in higher dimensionality is expected to be possible because the temporal coarse graining does not depend explicitly on the dimensionality of the system and on the HMF modelization for the self-electrostatic interactions at hand. This, in principle, opens to the possibility of including the contribution of a magnetic field along the curvilinear abscissa for the simple loop plasma model discussed here. We note that such additional ingredient is interesting not only for the coronal loops in the solar atmosphere but also in the context of fusion plasmas confined inside a Tokamak machine (see \citealt{goedbloed_keppens_poedts_2010,2018CoPP...58..457C}).\\
\indent The formalism of temporal coarse graining could be also useful to exospheric models of heliospheric plasmas  (i.e. the plasma that populates the interplanetary space) that describe interplanetary plasma as a non-collisional medium in contact with a fixed distribution at its boundary (\citealt{Chamberlain1960,Jockers1970,Lemaire1971,Lamy2003,Maksimovic1997,Zouganelis2004}). Such systems have the boundary at the base of the heliosphere (outer zone of the solar atmosphere), while the theoretical formalism discussed here allows one to implement an exospheric model that has the base in the high Chromosphere and that is able to reproduce the plasma of the solar atmosphere together with the heliospheric plasma. This approach can potentially give a model and a mechanism to produce suprathermal electrons in the heliospheric plasma, whose presence is suggested by in situ measurements (see for example \citealt{Pilipp_al_1987a,Halekas_al_2020,Maksimovic_al_2020,Berčič_2020}).\\
\indent Finally our formalism could also be used in the context of laser plasma interaction, where a region of the system is in thermal contact with another where the temperature fluctuates as a consequence of the external laser pumping (see e.g. \citealt{Gibbon_1996,Atzeni2001}).
\begin{acknowledgments}
We wish to thank Gianna~Cauzzi for very useful discussions. We acknowledge partial financial support from the MIUR-PRIN2017 project \textit{Coarse-grained description for non-equilibrium systems and transport phenomena (CO-NEST)} n.\ 201798CZL, the National Recovery and Resilience Plan, Mission 4 Component 2 - Investment 1.4 - NATIONAL CENTER FOR HPC, BIG DATA AND QUANTUM COMPUTING - funded by the European Union - NextGenerationEU - CUP B83C22002830001, the Solar Orbiter/Metis program supported by the Italian Space Agency (ASI) under the contracts to the National Institute of Astrophysics (INAF), Agreement ASI-INAF N.2018-30-HH.0, and the Fondazione CR Firenze under the projects \textit{HIPERCRHEL} and \textit{THE SWITCH}.
This research was partially funded by the European Union - Next Generation EU - National Recovery and Resilience Plan (NRRP) - M4C2 Investment 1.4 - Research Programme  CN00000013 "National Centre for HPC, Big Data and Quantum Computing" - CUP B83C22002830001 and by the European Union - Next Generation EU - National Recovery and Resilience Plan (NRRP)- M4C2 Investment 1.1- PRIN 2022 (D.D. 104 del 2/2/2022) - Project `` Modeling Interplanetary Coronal Mass Ejections'', MUR code 31. 2022M5TKR2,  CUP B53D23004860006.
Views and opinions expressed are however those of the author(s) only and do not necessarily reflect those of the European Union or the European Commission. Neither the European Union nor the European Commission can be held responsible for them.
\end{acknowledgments}
\bibliographystyle{jpp}
\bibliography{jpp-instructions}
\end{document}